\documentclass[10pt]{article} 
\usepackage[preprint]{tmlr}

\usepackage{amsmath,amsfonts,bm}

\def\eqref#1{equation~\ref{#1}}

\def\1{\bm{1}}

\DeclareMathAlphabet{\mathsfit}{\encodingdefault}{\sfdefault}{m}{sl}
\SetMathAlphabet{\mathsfit}{bold}{\encodingdefault}{\sfdefault}{bx}{n}

\usepackage{graphicx}
\usepackage{hyperref}
\usepackage{url}
\usepackage{algorithm}
\usepackage{algorithmic}
\usepackage{amsmath}
\usepackage{amssymb}
\usepackage{amsfonts}
\usepackage{amsthm}
\usepackage{multirow}
\usepackage{multicol}
\usepackage{booktabs}
\usepackage{array}

\newtheorem{theorem}{Theorem}
\newtheorem{lemma}{Lemma}
\newtheorem{definition}{Definition}
\newtheorem{proposition}{Proposition}

\newtheorem{remark}{Remark}

\newcommand{\FixMRV}{\mathrm{Fix}_{\mathrm{MRV}}}
\newcommand{\VarMRV}{\mathrm{Var}_{\mathrm{MRV}}}
\newcommand{\FixT}{\mathrm{Fix}_{T}}
\newcommand{\VarT}{\mathrm{Var}_{T}}
\newcommand{\CCP}{\mathrm{CCP}}

\title{A Cooperative Game Theoretic Model for \\ Sustainable, Smallholder  Carbon Farming}

\author{\name Geetha Charan \email geethacharan@iisc.ac.in \\
    \name Thandava Sunkara \email thandavas@iisc.ac.in\\
    \name Rohit P Suresh \email rohitpsuresh@iisc.ac.in \\
    \name V. Priyanka \email priyankav@iisc.ac.in \\
    \name Manojkumar Patil \email pmanojkumar@iisc.ac.in \\
    \name Kodur Sai Vinay Sathvik \email sathvik2004@gmail.com \\
    \name Y. Narahari \email narahari@iisc.ac.in \\
    \addr Indian Institute of Science
}

\def\month{MM}  
\def\year{YYYY} 
\def\openreview{\url{https://openreview.net/forum?id=XXXX}} 

\begin{document}

\maketitle

\begin{abstract}
Carbon farming offers a promising path to improve farmer livelihoods, mitigate climate change, and make farming sustainable. However, smallholder farmers, who dominate agriculture in developing economies, face structural barriers to participating in carbon farming due to high certification costs, limited landholdings, and limited budgets available to them. In this paper, we develop a cooperative game-theoretic framework to study how coalition formation can enable viable and profitable participation of smallholder farmers in carbon farming. Our approach explores three components: (i) Solo Portfolio Optimization (SPO), which determines an optimal portfolio of carbon farming practices that individual farmers can adopt under budget constraints; (ii) Joint Portfolio Optimization (JPO), which computes the maximum total economic surplus achievable through the formation of farmer coalitions; and (iii) surplus allocation among participating farmers in a principled way using cooperative game-theoretic solution concepts, such as Shapley value, core, least core, and nucleolus. We show that the underlying cooperative game is superadditive under an affordability condition and non-convex in general. As part of this study, we release \textsc{CarbonFARM-Bench}, a carbon farming dataset constructed by experts, and conduct controlled computational experiments on this dataset. We observe that solo participation in carbon farming is largely infeasible for smallholder farmers under baseline parameters, whereas coalition formation creates substantial value driven by economies of scale. In addition, we demonstrate that stability-aware allocation mechanisms are required for sustaining cooperation. Together, our results highlight that enabling smallholder carbon farming is fundamentally a challenge of institutional design, and we offer several policy recommendations to make it attractive to smallholder farmers.
\end{abstract}

\section{Introduction}
Agriculture forms the economic backbone of many developing economies, supporting the livelihoods of a large share of the population. However, smallholder farmers, who typically cultivate fragmented landholdings (of less than 2 hectares), operate with limited capital and often face distress. 
Erratic rainfall, volatile commodity prices, and rising input costs translate directly into income shocks, leaving small and marginal farmers perpetually exposed to distress.
At the same time, agriculture accounts for approximately 12\% of global greenhouse gas (GHG) emissions \citep{climatewatch2025}, placing the sector at the center of a fundamental tension: how to sustain and enhance agricultural productivity while meaningfully reducing its carbon footprint.

Against this backdrop, carbon farming has emerged as a promising path towards simultaneously addressing sustainability and livelihood concerns \citep{sarkar2025carbon}.  It encompasses crop cultivation practices such as conservation tillage, reduced fertilizer and pesticide application, improved irrigation, and need-based organic or bio-fertilization, that strengthen the capacity of agricultural systems to sequester atmospheric CO$_2$ or reduce net GHG emissions \citep{srivastava2024advancements}. The benefits extend well beyond climate mitigation: these practices can lower input costs, enrich agro-biodiversity, and improve water and nutrient efficiency, thus generating tangible co-benefits for the farmers who adopt them \citep{yadav2025carbon}. Crucially, verified emission reductions and removals can be certified and traded as carbon credits in voluntary carbon markets, thereby opening an additional, potentially stable revenue stream for smallholders \citep{priyanka2026carbon}.

However, realizing this potential is not straightforward for smallholder farmers. Carbon credit programs incur substantial fixed costs, such as measurement, reporting, and verification (MRV), third-party auditing, and registry fees, that are often independent of farm size \citep{vanderheyden2026monitoring}. For a farmer cultivating one or two hectares, these costs can easily eclipse the value of carbon credits generated, making solo participation economically non-viable \citep{shames2013reducing}. Coalition formation emerges as a natural institutional alternative. By pooling land and coordinating practices, groups of smallholders can collectively generate credit volumes that justify shared certification costs and present a credible counterparty to buyers and registries.

Applying cooperative action to carbon farming, however, introduces many challenges. The fundamental tension lies between the heterogeneity of carbon farming practices and the stability of any coalition formed around them. Even within a single village, smallholders vary considerably in farm size, soil quality, and access to capital. As a result, the portfolio of practices that maximizes net profit for one farmer may or may not be workable for another farmer. Moreover, even when two farmers adopt the same portfolios of practices, their actual sequestration outputs might differ. So, within a coalition, the members will rationally pursue different practice portfolios, and these individual choices jointly determine both the total carbon revenue the coalition can earn and the structure of the shared MRV costs, which scale with aggregate acreage. Crucially, in practice farmers do not pool their own input costs, yield outcomes, or finances: these remain private, and farmers are unwilling to bear another's agronomic costs. This raises a problem that goes beyond the question of whether a coalition is collectively worthwhile. For a coalition to be viable, its \emph{shared surplus} must be distributed so that, after each farmer's private costs and benefits are accounted for, every subgroup has a compelling reason to stay rather than go solo or form a rival group.

In this paper, we model each farmer's optimal carbon-practice portfolio as a constrained optimization problem, which we call the solo portfolio optimization problem. Next, we model coalition formation as a cooperative game in which the value of any coalition is the maximum total economic surplus (pooled carbon surplus plus members' private gains) achievable through joint portfolio optimization, subject to individual feasibility constraints and cost-sharing rules. Finally, we model surplus allocation as a linear program over the pooled carbon surplus, comparing core-based allocations (core, least core, and the nucleolus) against the Shapley value and standard baselines.

\subsection{Contributions}
This paper makes the following contributions.
\begin{itemize}
\item \textbf{A Cooperative Game-Theoretic Framework for Carbon Farming.}
We develop a three-phase framework that integrates individual decision-making, coalition formation, and revenue allocation. Through this, we provide a principled approach to modeling and exploring smallholder participation in carbon farming and carbon markets.

\item \textbf{CarbonFARM-Bench Dataset.} We release \textsc{CarbonFARM-Bench} (\textbf{F}air \textbf{A}llocation and \textbf{R}evenue \textbf{M}aximization for Smallholders in Carbon Farming Benchmark), an expert-constructed, literature-informed benchmark for smallholder  carbon farming, covering eight practices, pairwise interaction matrices, MRV and transaction costs, and synthetic farmer panels, to support reproducible research on coalition formation and surplus allocation.

\item \textbf{Structural Characterization of Coalition Game.}
We formally prove that the underlying coalition game is superadditive under an affordability condition, and non-convex in general.
\item \textbf{Evaluation of Revenue Allocation Mechanisms.} 
We systematically compare six revenue allocation mechanisms: four based on solution concepts in cooperative game theory (core, least core, nucleolus, Shapley value) and two standard baselines (area-proportional allocation and attributable-revenue allocation). Our results show that stability-driven approaches are required to ensure participation and prevent coalition deviations.
\item \textbf{Policy-Relevant Insights for Inclusion of Smallholder Farmers.} 
Based on empirical observations, we provide policy recommendations that help smallholder farmers access the benefits of carbon farming. 
\end{itemize}

\section{Review of Relevant Work}

\paragraph{Smallholder Carbon Farming Economics.}
A longstanding literature examines the costs that make carbon projects difficult for smallholders to implement.
\citet{cacho2005transaction} provide a typology of transaction costs for carbon-sink projects in developing countries. In contrast, \citet{cacho2007abatement} derive a feasibility frontier that identifies the minimum viable project size at a given carbon price.
\citet{cacho2013transaction} extend this analysis to offset exchange between a project developer and a group of landholders. However, by treating the group as a single seller, they leave the distribution of internal surplus unmodeled.
\citet{shames2013reducing} discuss ways to reduce costs and improve benefits to support the scaling up of smallholder carbon projects.
More recent work focuses on monitoring, reporting, and verification (MRV).
\citet{smith2020measure} review soil-carbon MRV approaches and propose a global MRV framework, highlighting why MRV costs can be high and partly fixed.
\citet{oldfield2022crediting} emphasize the trade-off between measurement rigor and cost in soil-carbon crediting.
\citet{vanderheyden2026monitoring} develop a cost model for four MRV systems and find that costs vary substantially across monitoring approaches and project designs, suggesting that policy support may be needed to make participation viable for smaller farms.

Field evidence further illustrates how these barriers can translate into unequal participation and benefits.
\citet{cariappa2025carbon} surveyed 841 farmers in seven carbon-farming villages in Haryana and Madhya Pradesh, India.
They find that participants were predominantly large landholders, only 4\% were women, and about 99\% had received no payment.
Moreover, 28\% had already discontinued the practices, indicating that participation does not necessarily translate into sustained adoption.
\citet{aggarwal2020reducing} find that participation in a forest-carbon project in India involved foregone crop, fodder, and fuelwood benefits, with small and marginal farmers particularly affected.
Together, these findings suggest that carbon farming's distributional outcomes depend not only on potential carbon revenue but also on participation, payment delivery, and the opportunity costs borne by farmers.

On the agronomic side, \citet{li2013carbon} evaluate paddy-field practices in terms of their carbon sequestration capacity and contribution to household income, while \citet{piscitelli2023carbon} develop bottom-up approaches to identify strategies for increasing soil carbon stocks.
\citet{pham2025stacking} show that combining practices can yield greater abatement and profit than adopting any single practice, highlighting the potential value of considering practice portfolios rather than isolated interventions.
Our Solo and Joint Portfolio Optimization problems build on this perspective by incorporating farmer-level budget constraints and shared certification costs.
Taken together, these studies establish the economic and agronomic rationale for collective carbon farming, but leave open how the surplus generated by a pooled project should be distributed among heterogeneous farmers.

\paragraph{Cooperatives and Collective Action.}
\citet{markelova2009collective} review how collective organization helps smallholders reduce transaction costs and access markets, while \citet{ostrom1990governing} document how groups can sustain cooperation without external enforcement.
However, the benefits of collective action are not uniform across settings.
\citet{agarwal2018group} find that group farms in Kerala outperformed individual family farms in output per hectare and net returns, whereas those in Telangana performed worse.
\citet{bikkina2018fpo} show that farmer producer organizations (FPOs) in Gujarat can lower transaction costs through collective action, but remain constrained by limited capital.
Participation is also heterogeneous: a meta-analysis of 168 studies finds that farm size, access to credit, and distance to markets are significantly associated with cooperative membership, although the estimated effect sizes are generally small \citep{grashuis2025determinants}.
Together, these findings suggest that collective organization can create economic benefits, but its effectiveness and accessibility depend on local conditions and farmers' resources.

Cooperatives have also been designed to address specific agricultural activities.
\citet{bhardwaj2023designing} use farmer collectives to design fair and cost-optimal auctions for procuring agricultural inputs.
\citet{sarkar2023coalition} study agricultural cooperatives using multi-agent coalitional models and show that similarity-based coalition formation can generate positive revenue for smallholders.
\citet{zheng2022cooperative} develop cooperative game-based incentive mechanisms to support biopesticide adoption through FPOs.
In the carbon setting, empirical studies associate cooperatives with lower agricultural emissions, mainly through reduced fertilizer use \citep{wang2024effects,zhang2025can}, while \citet{mantey2025carbon} examine the role of dairy cooperatives in smallholder carbon projects.
Although these studies highlight the potential of cooperation for agricultural and environmental outcomes, they do not examine how the resulting carbon surplus is distributed among participating farmers.
In practice, carbon projects are also increasingly moving toward aggregation.
The VCS methodology VM0042 supports grouped projects \citep{verra_vm0042}, and an AWD-based project in Pakistan pooled smallholders under a single certification umbrella to obtain the first rice methane credits under the Gold Standard \citep{netzeroag_rice_methane_credits}.
These developments illustrate how aggregation can facilitate certification, while leaving the question of internal surplus allocation open.

\paragraph{Incentives and Collective Payments.}
\citet{raina2024incentive} map incentive mechanisms in carbon-farming contracts and distinguish results-based, action-based, and hybrid payments, highlighting how payment design, monitoring, and risk allocation shape farmer participation.
\citet{suich2013effectiveness} find in Mozambique and Namibia that incentives may fail to sustain participation when they are poorly communicated, unsuitable for participants, or inadequately delivered.
\citet{chu2025evolutionary} analyze a tripartite evolutionary game among government, farmer cooperatives, and farmers under carbon trading, finding that a higher share of carbon revenue for farmers increases participation.
These studies highlight the importance of payment design for participation, but \citet{chu2025evolutionary} assume a fixed sharing ratio within a non-cooperative framework, whereas our approach derives the sharing rule from coalition-based allocation.

Collective environmental payments have also been examined in the agri-environment literature.
The agglomeration bonus in \citet{parkhurst2002agglomeration} provides landowners with an additional payment when enrolled parcels are spatially connected, thereby rewarding coordinated participation.
\citet{kuhfuss2016nudging} show in a choice experiment that a bonus conditional on a group enrollment threshold increases farmers' willingness to join.
\citet{watzold2014agglomeration} find that the relative performance of homogeneous payments, agglomeration bonuses, and agglomeration payments depends on whether participants can make side payments. This finding is relevant to our transferable-utility assumption, under which coalition members can redistribute surplus among themselves.

\paragraph{Cooperative Games and Allocation.}
Cooperative game theory provides a framework for analyzing profit allocation in agriculture and natural-resource management.
\citet{bekolli2025profit} study multidistributor-farmer games in which farmers sell through a cooperative, establish the existence of stable allocations, and characterize a rule that compensates farmers.
\citet{bekolli2025harvesttech} introduce totally balanced games for firms sharing crop-processing technology and derive core-stable compensation rules.
\citet{parrachino2006cooperative} review cooperative games in environmental and water-resource problems, while \citet{ambec2002sharing} provide a canonical analysis of river sharing among heterogeneous agents, characterizing a core-stable distribution.
Together, these studies illustrate how cooperative game theory can be used to analyze surplus distribution and allocation stability.

\citet{frisk2010cost} allocate the savings from pooled transport planning among eight Swedish forest companies. They compute coalition values using a linear program and compare the Shapley value, nucleolus, separable costs, shadow prices, and volume weights.
\citet{guajardo2016review} survey allocation rules in collaborative transportation, including how frequently these rules yield core allocations.
In linear production games, coalition values are obtained from linear programs, and LP duality guarantees a core allocation under the model's assumptions \citep{owen1975core}.
By contrast, \citet{caprara2010new} study integer minimization games in which coalition values are defined by integer programs, and compute cost shares using row and column generation.
\citet{petrosjan2003time} extend cooperative allocation to dynamic settings by developing a time-consistent Shapley allocation for joint emission reduction.

\paragraph{Solution Concepts and Computation.}
The core \citep{gillies1959solutions}, least core \citep{maschler1979geometric}, nucleolus \citep{schmeidler1969nucleolus}, and Shapley value \citep{shapley1953value} are the standard tools for allocating the surplus of a coalition.
Non-emptiness of the core is characterized by balancedness \citep{bondareva1963,shapley1967balanced}, and convex games have a non-empty core containing the Shapley value \citep{shapley1971cores}.
Exact computation requires evaluating all $2^n$ coalitions, each of which is an optimization problem in our setting, so cost grows quickly with the number of farmers.
The nucleolus can be computed by a sequence of nested linear programs \citep{kopelowitz1967}, and \citet{benedek2021finding} give a faster descent algorithm with a verification test.
\citet{yan2021if} show that arbitrarily good approximations of the least core can be computed efficiently, and they make a case for the least core as an alternative to the Shapley value in credit assignment, while proving an impossibility result for the nucleolus.
For larger coalitions, sampling-based approximations exist for the Shapley value \citep{mitchell2022sampling}, and core-like stability can be guaranteed with high probability from sampled coalition values \citep{balcan2015learning}.

\paragraph{Gap.}
Existing studies either examine the cost barriers to smallholder carbon farming without addressing how the surplus from collective participation is distributed, or investigate cooperative allocation in agriculture without incorporating the carbon-specific features of practice portfolios, farmer-level budget constraints, heterogeneous agronomic payoffs, and economies of scale in certification. The interaction between individual practice choices, the gains from collective certification, and the allocation of coalition surplus therefore remains insufficiently explored. This study addresses this gap by integrating Solo and Joint Portfolio Optimization with cooperative game-theoretic allocation rules.

\section{Preliminaries}
\label{section:prelims}

\subsection{TU Game}
A cooperative game with transferable utilities (TU) is defined by a pair $(N,v)$, where $N=\{1,2,\dots,n\}$ is the set of players and $v:2^N \to \mathbb{R}$ is the characteristic function satisfying $v(\emptyset)=0$. For every coalition $S \subseteq N$, the quantity $v(S)$ denotes the total utility that the members of $S$ can jointly generate and redistribute among themselves without loss.

In our setting, the players are farmers, and the characteristic function $v(S)$ represents the maximum jointly attainable economic surplus of the coalition $S$ subject to the relevant constraints. Once the coalition values are determined, the key question becomes how to distribute the pooled carbon surplus among the farmers in a way that is fair, interpretable, and resistant to defection. We address this through the cooperative game-theoretic solution concepts introduced in the following sections.

\subsection{The Core}
\label{subsec:core}
Let $\mathbf{x}=(x_i)_{i\in N}$ be an allocation vector, where $x_i$ denotes the payoff assigned to player $i$. The core \citep{scarf1967core} is the set of all allocations satisfying
\begin{equation}
\sum_{i\in N} x_i = v(N)
\end{equation}
\begin{equation}
\text{and} \quad \sum_{i\in S} x_i \ge v(S), \qquad \forall S \subseteq N.
\end{equation}
The first condition is \emph{efficiency}: the entire value of the grand coalition is distributed among the players. The second condition is \emph{coalitional rationality}: every coalition receives at least as much as it could secure on its own. Hence, an allocation in the core is stable against deviations by any subgroup of players.

\subsection{The Least Core}
\label{subsec:least_core}
When the core is empty, the coalitional stability constraints can be relaxed by allowing a uniform slack $\varepsilon \ge 0$. The least core \citep{maschler1979geometric} is defined through the optimization problem
\begin{equation}
\begin{gathered}
\min_{\mathbf{x},\varepsilon} \ \varepsilon \\
\text{s.t.} \quad \sum_{i\in N} x_i = v(N) , \\ 
\qquad
\sum_{i\in S} x_i \ge v(S)-\varepsilon \ \ \forall S \subseteq N.
\end{gathered}
\end{equation}
The parameter $\varepsilon$ measures the maximum amount by which any coalition's value claim may be violated. Thus, the least core identifies allocations that minimize the largest dissatisfaction across all coalitions.

The least core is especially useful when exact core stability is unattainable. Rather than discarding stability altogether, it finds the ''closest'' possible approximation to the core.

\subsection{The Nucleolus}
\label{subsec:nuc}
For an allocation $\mathbf{x}$, define the excess of coalition $S$ as
\begin{equation}
e(S,\mathbf{x}) = v(S) - \sum_{i\in S} x_i.
\end{equation}
The excess measures the dissatisfaction of coalition $S$ under allocation $\mathbf{x}$: a larger value of $e(S,\mathbf{x})$ means that coalition $S$ is further from receiving its worth. The nucleolus \citep{schmeidler1969nucleolus} is the unique allocation that lexicographically minimizes the ordered vector of coalition excesses, beginning with the largest excess, then the second largest, and so on.

Intuitively, the nucleolus refines the least core by not only minimizing the maximum dissatisfaction, but also resolving remaining conflicts as evenly as possible across coalitions. In games with a non-empty core, the nucleolus always belongs to the core.

\subsection{The Shapley Value}
\label{subsec:shap}
The Shapley value \citep{shapley1953value} allocates the worth of the grand coalition according to each player's expected marginal contribution across all possible coalition formation orders. Formally, for each player $i\in N$, Shapley value is defined as 
\begin{equation}
\phi_i(v)=\sum_{S \subseteq N\setminus\{i\}}
\frac{|S|!(|N|-|S|-1)!}{|N|!}
\left[v(S\cup\{i\})-v(S)\right].
\end{equation}

The Shapley value is widely used as it satisfies several desirable axioms, including efficiency, symmetry, dummy-player treatment, and additivity. It therefore provides a natural fairness benchmark based on contribution. However, unlike the core, it is not defined in terms of coalitional stability, and a Shapley allocation need not belong to the core. In applications such as coalition formation among farmers, the Shapley value is therefore best viewed as a marginal-contribution benchmark rather than a guaranteed stable allocation rule.

\section{Methodology}
\label{sec:method}

\subsection{Notation}
Let $F=\{1,\dots,n\}$ be the set of farmers and $P=\{1,\dots,m\}$ the set of carbon-farming practices.
For farmer $F_i$ and practice $P_j$ we use:
\begin{itemize}
  \item $\mathrm{FS}_i$: farm size of $F_i$ (ha);
  \item $B_i$: budget available to $F_i$ (\$);
  \item $\mathrm{OC}_j$: operational cost of $P_j$ relative to baseline (\$/ha);
  \item $\mathrm{CMP}_{ij}$: farmer-specific carbon mitigation potential of $P_j$ on $F_i$'s land (tCO$_2$e/ha);
  \item $Y_j$: percentage yield change of $P_j$ relative to conventional cultivation;
  \item $Y^{0}$: conventional yield (t/ha);
  \item $\mathrm{CP}$: crop price (\$/t);
  \item $\mathrm{CCP}$: carbon credit price (\$/tCO$_2$e).
\end{itemize}
Due to variations in soil composition, geography, topography, and agro-climatic conditions, we assume CMP varies across farmers. Whereas operational costs and yield impacts are assumed to be uniform across farmers. This is a modeling simplification; the framework can be extended to farmer-specific $\mathrm{OC}_{ij}$ and $Y_{ij}$ without altering the model's overall structure. A complete notation list is provided in Appendix~\ref{appendix:notation_summary}, though each symbol is also introduced in the text when it first appears.

We emphasize that CMP in this study is an economic modeling quantity representing potential CO2-equivalent mitigation, rather than a claim that this amount would qualify as tradable carbon credits under a particular registry or methodology. It should be noted that project-level additionality, permanence, leakage, and uncertainty deductions, as well as registry-specific eligibility requirements, are outside the scope of the present model.

\subsection{Practice Portfolios and Interaction Effects}
A \emph{portfolio} is a subset of practices ($\Pi\subseteq P$) that a particular farmer takes up for carbon farming.
Let $I\subseteq\{(j,k):j<k\}$ be the set of hard-incompatible practice pairs that represent combinations of practices that cannot be done together on the field. Therefore, a portfolio is \emph{admissible} if it contains no pair from $I$.

Joint adoption of practices can produce biological and agronomic interactions that are not captured by summing individual effects \citep{billick1994higher}. For example, agroforestry and cover cropping may jointly sequester more carbon than the sum of their separate contributions (synergy), and two reduced-tillage practices may share machinery costs (cost complementarity). We model these effects pairwise, using symmetric parameters $\alpha_{jk}=\alpha_{kj}$, $\beta_{jk}=\beta_{kj}$, and $\gamma_{jk}=\gamma_{kj}$ for each pair $(P_j,P_k)$:
\begin{itemize}
    \item $\alpha_{jk}$: synergistic or antagonistic effect on sequestration when $P_j$ and $P_k$ are jointly adopted. $\alpha_{jk} > 0$ indicates a sequestration synergy and $\alpha_{jk} < 0$ indicates an antagonism.
    \item $\beta_{jk}$: cost interaction. $\beta_{jk} < 0$ captures shared infrastructure or input savings and $\beta_{jk} > 0$ captures additional management complexity.
    \item $\gamma_{jk}$: yield interaction. $\gamma_{jk} > 0$ indicates a compounding yield benefit from the combination and $\gamma_{jk} < 0$ a yield penalty.
\end{itemize}

A portfolio of size $|\Pi|$ contains $|\Pi|(|\Pi|-1)/2$ pairs. To keep the total interaction from growing with portfolio size \citep{bairey2016high}, we scale the pairwise terms by the interaction weight
\begin{equation}
  w(\Pi)=
  \begin{cases}
    \dfrac{2}{|\Pi|(|\Pi|-1)}, & |\Pi|\ge 2,\\[2ex]
    0, & |\Pi|\le 1,
  \end{cases}
  \label{eq:w}
\end{equation}
so that the total interaction is the \emph{average} of the pairwise effects. The portfolio-level carbon mitigation, operational cost, and yield revenue change are then
\begin{align}
  \mathrm{CMP}_i(\Pi) &= \sum_{j\in\Pi}\mathrm{CMP}_{ij}
      + w(\Pi)\sum_{\substack{j<k\\ j,k\in\Pi}}\alpha_{jk}, \label{eq:cmp}\\[1.5ex]
  \mathrm{OC}(\Pi) &= \sum_{j\in\Pi}\mathrm{OC}_{j}
      + w(\Pi)\sum_{\substack{j<k\\ j,k\in\Pi}}\beta_{jk}, \label{eq:oc}\\[1.5ex]
  \Delta Y(\Pi) &= \frac{Y^{0}\,\mathrm{CP}}{100}
      \Biggl(\sum_{j\in\Pi}Y_j
      + w(\Pi)\sum_{\substack{j<k\\ j,k\in\Pi}}\gamma_{jk}\Biggr). \label{eq:dy}
\end{align}
Here $\Delta Y(\Pi)$ is the yield revenue change (\$/ha), positive for a gain and negative for a loss, and $\mathrm{OC}(\Pi)$ may be negative (a saving).

Reliable field estimates of higher-order practice interactions are scarce. The parameters $\alpha_{jk}$, $\beta_{jk}$, and $\gamma_{jk}$ should therefore be read as scenario parameters, not as universally estimated agronomic effects.

\subsection{Certification Costs}
Carbon credit certification involves two categories of costs: Monitoring, Reporting, and Verification (MRV) costs and Transaction costs (T). For a coalition $S$ with total area $A(S)=\sum_{i\in S}\mathrm{FS}_i$, we define
\begin{align}
  C_{\mathrm{MRV}}(S) &= \mathrm{Fix}_{\mathrm{MRV}} + \mathrm{Var}_{\mathrm{MRV}}\,A(S)^{\delta},
      \quad \delta\in(0,1), \label{eq:cmrv}\\
  C_{T}(S) &= \mathrm{Fix}_{T} + \mathrm{Var}_{T}\,|S|, \label{eq:ct}\\
  C(S) &= C_{\mathrm{MRV}}(S) + C_{T}(S). \label{eq:cs}
\end{align}
$\mathrm{Fix}_{\mathrm{MRV}}$ (\$) covers the base administrative and auditing overhead incurred one time per certified unit, regardless of size, $\mathrm{Var}_{\mathrm{MRV}}$ (\$/ha$^{\delta}$) scales with certified area, and $\delta\in(0,1)$ is a sublinearity parameter capturing economies of scale in monitoring \citep{logan2011economies}. Because $\delta<1$, the MRV cost is concave in total area, so pooling farms under a single certification umbrella strictly reduces per-hectare MRV costs relative to individual certification.
$\mathrm{Fix}_{T}$ (\$) is a one-time registry and contracting overhead, and $\mathrm{Var}_{T}$ (\$/farmer) is a per-farmer administration fee \citep{ducos2009agri}. For a singleton,
\[
  C(\{i\})=\mathrm{Fix}_{\mathrm{MRV}}+\mathrm{Var}_{\mathrm{MRV}}\,\mathrm{FS}_i^{\delta}
  +\mathrm{Fix}_{T}+\mathrm{Var}_{T}.
\]

\subsection{Shared and Private Quantities}
\label{sec:shared-private}
Farmers in a coalition file jointly and generate credits together, but they do not pool agronomic costs or outcomes.
No farmer bears another's operational costs or yield losses, and none shares another's yield gains.
We therefore partition the economics into two groups.

\paragraph{Shared (pooled) quantities.}
The carbon revenue and the certification cost,
\begin{equation}
  \begin{aligned}
    R(S,\Pi^S) &= \sum_{i\in S} \mathrm{FS}_i\cdot\mathrm{CMP}_i(\Pi)\cdot\mathrm{CCP},\\
    C(S)       &= C_{\mathrm{MRV}}(S)+C_{T}(S).
  \end{aligned}
  \label{eq:shared}
\end{equation}
Here, $R(S,\Pi^S)$ is the coalition's pooled revenue, where each farmer $ F_i$'s revenue is the sequestration per hectare, scaled by farm size and the prevailing carbon credit price. $C(S)$ is the joint certification cost from \eqref{eq:cs}, which arises only from filing together and does not depend on any farmer's chosen portfolio. These two are the only quantities that are pooled; everything else is private to each farmer.

\paragraph{Private quantities.}
Operational cost and yield effects, which belong to each farmer alone:
\begin{equation}
  g_i(\Pi)=\mathrm{FS}_i\bigl(\Delta Y(\Pi)-\mathrm{OC}(\Pi)\bigr).
  \label{eq:g}
\end{equation}
The term $\mathrm{FS}_i\,\Delta Y(\Pi)$ is the revenue from yield gain (or loss), positive when practices improve yield and negative when they reduce it, and $\mathrm{FS}_i\,\mathrm{OC}(\Pi)$ is the operational cost. Because $\Delta Y(\Pi)$ and $\mathrm{OC}(\Pi)$ can each change sign, they jointly govern the net agronomic payoff $g_i(\Pi)$: a yield loss enters as a phantom cost, while a cost-reducing practice enters as a phantom revenue.


\paragraph{Two surpluses.}
Throughout the paper we distinguish two quantities for a coalition $S$ with selected profile $\Pi^S$:
\begin{align}
  p(S) &= R(S,\Pi^S)-C(S) &&\text{(pooled, transferable carbon-market surplus)},\label{eq:p}\\
  G(S) &= \sum_{i\in S}g_i(\Pi^S_i) &&\text{(private, non-transferable gains)},\label{eq:G}\\
  v(S) &= p(S)+G(S) &&\text{(total economic surplus).}\label{eq:vsplit}
\end{align}
Only $p(S)$ can be moved between farmers. The characteristic function $v(S)$ defined measures total economic surplus, because the coalition selects its portfolios to maximize what its members actually earn.

\paragraph{Non-pooling premise.}
Non-pooling means that private costs and yields are never transferred, never charged to another farmer, and never used to fund another farmer's participation. It does \emph{not} mean that private effects are invisible to portfolio selection.
We encode it through three rules:
\begin{description}
  \item[(NP1) Objective.] Portfolios are selected to maximize total economic surplus. Private quantities enter the objective as welfare, but only through each farmer's own portfolio.
  \item[(NP2) Constraints.] Each farmer's private economics constrain that farmer's own portfolio through a farmer-specific constraint, never through an aggregate one.
  \item[(NP3) Transfers.] Only the pooled surplus $p(F)$ is distributed. A farmer's final surplus is
  \begin{equation}
    u_i = x_i + g_i(\Pi_i),
    \label{eq:u}
  \end{equation}
  where $x_i$ is her share of the pooled surplus and $g_i$ is retained privately.
\end{description}
This is a quasi-linear utility structure: only $x_i$ is transferable, while $g_i(\Pi_i)$ stays with the farmer. It formalizes our assumption that farmers do not pool their input costs or yield outcomes, but still care about them.

\paragraph{Per-farmer constraints.}
Let $S$ be a coalition and let
\[
  \kappa_i(S)=\frac{\mathrm{FS}_i}{A(S)}\,C(S)
\]
be the area-proportional certification outlay of farmer $i$.
A portfolio $\Pi_i$ is \emph{individually admissible in $S$} if it is admissible with respect to $I$ and
\begin{align}
  \text{(Budget)}\qquad & B_i + g_i(\Pi_i)\ \ge\ \kappa_i(S), \label{eq:budget}\\
  \text{(Self-reliance)}\qquad & R_i(\Pi_i)+g_i(\Pi_i)-\kappa_i(S)\ \ge\ 0. \label{eq:selfrel}
\end{align}
The budget constraint requires that the farmer's funds, adjusted for her own yield and operating effects, cover her certification outlay.
Self-reliance requires that the benefit generated on her own land covers her own certification outlay.

\subsection{Solo Portfolio Optimization (SPO)}
A farmer $F_i$ acting alone either adopts a carbon-farming portfolio and certifies, or does not participate and earns nothing.
If she participates, she selects a portfolio $\Pi^{\mathrm{so}}_i\subseteq P$ by solving
\begin{align}
  \Pi^{\mathrm{so}}_i=\arg\max_{\Pi\ \text{admissible}}\ & R_i(\Pi)+g_i(\Pi)-C(\{i\}) \label{eq:spo}\\
  \text{s.t.}\ & B_i+g_i(\Pi)\ge C(\{i\}),\nonumber\\
  & R_i(\Pi)+g_i(\Pi) \ge C(\{i\}).\nonumber
\end{align}
\noindent The objective is the carbon revenue $R_i(\Pi)$ plus the private net agronomic payoff $g_i(\Pi)$, minus the certification cost $C(\{i\})$, which collects the monitoring, reporting, and verification (MRV) and transaction costs. The budget constraint ensures that net costs do not exceed the farmer's available capital $B_i$, and the self-reliance constraint ensures that the benefit generated on her own land covers her certification outlay.

\paragraph{Standalone value.}
Let $\mathrm{OPT}_i$ denote the optimal value of \eqref{eq:spo}, with $\mathrm{OPT}_i=-\infty$ if the problem is infeasible. The standalone value of farmer $F_i$, which is her total surplus from acting alone, is
\begin{equation}
  v(\{i\})=\max\{0,\mathrm{OPT}_i\}.
  \label{eq:vsolo}
\end{equation}
Solo participation is feasible when $\mathrm{OPT}_i>0$. Otherwise $F_i$ does not participate, $\Pi^{\mathrm{so}}_i=\emptyset$, and $v(\{i\})=0$.
The value $v(\{i\})$ is the surplus farmer $F_i$ retains after all costs are settled, and it serves as her outside-option payoff. Every allocation rule must guarantee each farmer at least $v(\{i\})$ for individual rationality to hold.

\subsection{Joint Portfolio Optimization (JPO)}
\label{sec:jpo}
When a coalition $S\subseteq F$ with $|S|\ge 2$ forms, its members file for certification jointly and share MRV and transaction costs $C(S)$.
Each farmer $F_i\in S$ selects a \textbf{coalition portfolio} $\Pi^S_i\subseteq P$. These portfolios may differ across farmers, and may differ from the farmer's solo portfolio $\Pi^{\mathrm{so}}_i$, because the reduced certification burden can make additional or different practices financially viable.

Operational costs and yield-related revenue are private to each farmer; only certification costs are pooled. Pooling these farmer-specific quantities would be unfair. Farms differ across farmers in size, soil, climate, and other local conditions, and portfolios differ as well, so the practices adopted, the operational costs incurred, and the yield gains or losses realized all vary from one farmer to another. A farmer who adopts a costly practice, who operates a farm with less favorable conditions, or who experiences a yield loss would otherwise have part of that burden shifted onto others. In contrast, a farmer with low costs or high yield gains would have to share the benefit. Certification costs, by contrast, arise from the joint filing itself and are not tied to any one farmer's choices or farm conditions so that they can be shared in proportion to area. Each farmer's own economics, therefore, enter the coalition's problem only through her portfolio and constraints \eqref{eq:budget}--\eqref{eq:selfrel}.

The coalition selects portfolios to maximize its members' total economic surplus:
\begin{align}
  \mathrm{OPT}(S)=\max_{\{\Pi_i\}_{i\in S}}\ & \sum_{i\in S}\bigl[R_i(\Pi_i)+g_i(\Pi_i)\bigr]-C(S) \label{eq:jpo}\\
  \text{s.t.}\ & \Pi_i\ \text{individually admissible in } S\quad \forall i\in S.\nonumber
\end{align}
There are $|S|$ pairs of constraints, one per farmer.
That is, each farmer's budget, adjusted for yield gains or losses and operational costs, must cover her area-proportional share of the coalition's certification costs, and the benefit generated on her own land must cover the same share.
For $|S|=1$, \eqref{eq:jpo} coincides with \eqref{eq:spo}, so SPO is the singleton case of JPO.

\paragraph{Selected profile.}
When several profiles attain the optimum, ties are broken in favor of the profile that maximizes
\[
  \min_{i\in S}\bigl[R_i(\Pi_i)+g_i(\Pi_i)-\kappa_i(S)\bigr],
\]
and any remaining tie by a fixed ordering of portfolios.
This makes the selected profile $\Pi^S=\{\Pi^S_i\}_{i\in S}$ unique, and favors the profile that leaves the worst-off member with the largest net benefit.
If $\mathrm{OPT}(S)\le 0$ or no admissible profile exists, $S$ does not certify jointly, and we set $\Pi^S_i=\emptyset$ for all $i\in S$.

\paragraph{Characteristic function.}
\begin{equation}
  v(S)=\max\{0,\mathrm{OPT}(S)\}\quad(|S|\ge 1),\qquad v(\emptyset)=0,
  \label{eq:v}
\end{equation}
so $v(S)\ge 0$ for all $S$ and no coalition has value $-\infty$.
The characteristic function measures the maximum \emph{total economic surplus} of $S$, namely $v(S)=p(S)+G(S)$ as in \eqref{eq:vsplit}.
Only the pooled component $p(S)$ is transferable (NP3).



\subsection{Revenue Allocation}
\label{sec:allocation}
Once the grand coalition is formed, the next question is how to divide its pooled surplus among the farmers. An allocation $x=(x_i)_{i\in F}$ distributes the pooled surplus of the grand coalition. We write $x_i=\rho_i-\kappa_i(F)$, where $\rho_i$ is the share of gross carbon revenue received by farmer $i$ and $\kappa_i(F)$ is the certification outlay she actually pays. The outlay used in the budget constraint \eqref{eq:budget} is therefore the payment that farmers actually make, and $x_i$ is the resulting net share.

Let $\Pi^F$ be the profile selected by JPO for the grand coalition. Then $\sum_{i}x_i=p(F)$. After the payout, each farmer adds her private term, so her final surplus is
\begin{equation}
  u_i=x_i+g_i(\Pi^F_i),\qquad \sum_{i\in F}u_i=p(F)+G(F)=v(F).
  \label{eq:ufinal}
\end{equation}
Only $x_i$ is transferable, and $g_i(\Pi^F_i)$ always stays with the farmer. The self-reliance test \eqref{eq:selfrel} uses the revenue generated on $F_i$'s own land, $R_i(\Pi_i)$, only as a screen for admissible portfolios.

\paragraph{Deviation behavior.}
A subgroup $T\subsetneq F$ may leave the grand coalition and file for certification on its own. It can choose any profile that is individually admissible in $T$, including $\emptyset$, which means not certifying. The members of $T$ can share the pooled surplus of their chosen profile freely among themselves, but they cannot transfer private gains. A deviating singleton $\{i\}$ has nobody to pool with and behaves as in SPO. A farmer who leaves can also remain a conventional farmer with no portfolio and no carbon revenue, which is why $v(\{i\})$ is floored at zero in \eqref{eq:vsolo}.

\paragraph{Deviation value.}
A farmer who stays in the grand coalition keeps the private gain $g_i(\Pi^F_i)$. A deviating group must therefore be compared against this gain rather than the raw coalition value. For $\emptyset\ne T\subseteq F$ we define
\begin{equation}
  \hat v(T)=v(T)-\sum_{i\in T}g_i(\Pi^F_i),\qquad \hat v(\emptyset)=0.
  \label{eq:vhat}
\end{equation}
Equivalently,
\[
  \hat v(T)=p(T)+\sum_{i\in T}\bigl[g_i(\Pi^T_i)-g_i(\Pi^F_i)\bigr].
\]
It is the amount of transferable surplus that $T$ must receive for deviation to be unattractive, given that its members keep their private gains under the grand-coalition profile if they stay. For the grand coalition, $\hat v(F)=v(F)-G(F)=p(F)$. For a singleton, $\hat v(\{i\})=v(\{i\})-g_i(\Pi^F_i)$, and when solo participation is infeasible, $v(\{i\})=0$, so $\hat v(\{i\})=-g_i(\Pi^F_i)$. If $\Pi^T_i=\Pi^F_i$ for every $i\in T$, then $\hat v(T)=p(T)$.

\begin{proposition}[Blocking condition]
\label{prop:block}
Let $x$ be an allocation with $\sum_i x_i=p(F)$ and let $T\subsetneq F$ be a deviating group as described above. There exists a profile and a split of the transferable surplus of $T$ that make every member of $T$ strictly better off than under $x$ if and only if
\[
  \hat v(T)>\sum_{i\in T}x_i.
\]
\end{proposition}

\begin{proof}
Under $x$, member $i$ obtains $u_i=x_i+g_i(\Pi^F_i)$. Suppose $T$ deviates to an admissible profile $\Pi$. Its members share the pooled surplus
\[
  p_T(\Pi)=\sum_{i\in T}R_i(\Pi_i)-C(T)
\]
through transfers $y_i$ with $\sum_{i\in T}y_i=p_T(\Pi)$, and member $i$ obtains $y_i+g_i(\Pi_i)$. For a singleton, $y_i=p_T(\Pi)$.

The $y_i$ can be transferred freely inside $T$, whereas the $g_i$ cannot. So every member can be made strictly better off under $\Pi$ if and only if the totals satisfy
\[
  p_T(\Pi)+\sum_{i\in T}g_i(\Pi_i)\ >\ \sum_{i\in T}u_i.
\]
The left-hand side is the total surplus of $T$ under $\Pi$, and its maximum over admissible profiles is $v(T)$ by \eqref{eq:jpo} and \eqref{eq:v}. Hence, a suitable profile and split exist if and only if
\[
  v(T)\ >\ \sum_{i\in T}\bigl[x_i+g_i(\Pi^F_i)\bigr],
\]
which is the same as $\hat v(T)>\sum_{i\in T}x_i$.

For sufficiency, take $\Pi=\Pi^T$ and set
\[
  y_i=x_i+g_i(\Pi^F_i)-g_i(\Pi^T_i)+\eta,
  \qquad
  \eta=\frac{v(T)-\sum_{i\in T}\bigl[x_i+g_i(\Pi^F_i)\bigr]}{|T|}>0.
\]
Then
\[
  \sum_{i\in T}y_i=v(T)-\sum_{i\in T}g_i(\Pi^T_i)=p(T),
\]
as required.
\end{proof}
Only the totals matter here. The transferable shares can be redistributed inside $T$, while the private gains cannot, so they enter the test as fixed offsets.

\subsubsection*{Feasibility Constraints}
An allocation is admissible if it satisfies two conditions. The first is \emph{efficiency}, which requires that the entire pooled carbon-market surplus is distributed and nothing is left unallocated:
\begin{equation}
  \sum_{i\in F}x_i=p(F)=\hat v(F). \label{eq:eff}
\end{equation}
The second is \emph{stability}, which ensures that no subgroup of farmers can be made better off, in terms of total surplus \eqref{eq:ufinal}, by leaving and forming an independent coalition $T$:
\begin{equation}
  \sum_{i\in T}x_i\ \ge\ \hat v(T)\qquad \forall\,\emptyset\ne T\subsetneq F. \label{eq:stab}
\end{equation}
By Proposition~\ref{prop:block}, \eqref{eq:stab} is exactly the no-blocking condition for the class of deviations described above. Individual rationality is the special case $|T|=1$, where the constraint reads $x_i\ge v(\{i\})-g_i(\Pi^F_i)$, or equivalently $u_i\ge v(\{i\})$.

\subsubsection{Cooperative Game-Theoretic Solutions}
\label{sec:cgt_solutions}
We apply the solution concepts from Section~\ref{section:prelims} to the farmer coalition game, replacing the coalition worths in the stability constraints with the deviation values $\hat v(T)$. Every mechanism satisfies efficiency \eqref{eq:eff}. When the core is non-empty, we look for allocations inside it. Otherwise, we fall back on approximate stability concepts.

\paragraph{Least core.}
We solve the LP from Section~\ref{subsec:least_core} using the deviation values $\hat v(T)$ from \eqref{eq:vhat}:
\begin{equation}
  \begin{aligned}
    \min_{x,\varepsilon}\ & \varepsilon\\
    \text{s.t.}\ & \sum_{i\in F}x_i=p(F),\\
    & \sum_{i\in T}x_i\ge \hat v(T)-\varepsilon\quad\forall\,\emptyset\ne T\subsetneq F.
  \end{aligned}
  \label{eq:leastcore}
\end{equation}
If the optimal slack $\varepsilon^*\le 0$, the core is non-empty. If $\varepsilon^*>0$, no allocation is fully stable, and $\varepsilon^*$ is the smallest uniform violation. The least core minimizes the maximum coalitional dissatisfaction through a single LP, but it need not be unique, since several allocations can achieve the same optimal slack $\varepsilon^*$. Among them, we select the allocation that also minimizes $\sum_i x_i^2$.

\paragraph{Nucleolus.}
The nucleolus refines the least core to a unique allocation by lexicographically minimizing the coalition excesses $e(T,x)=\hat v(T)-\sum_{i\in T}x_i$. It always lies within the least core, and when the core is non-empty, it also lies within the core. We compute it through a sequence of LPs. At each stage, the constraints that bind at the previous optimum are fixed, and $\varepsilon$ is minimized over the remaining coalitions until the allocation is unique.

\paragraph{Shapley value.}
We compute the Shapley value on the deviation game, with $\hat v(\emptyset)=0$:
\begin{equation}
  \phi_i(\hat v)=\sum_{T\subseteq F\setminus\{i\}}
  \frac{|T|!\,(|F|-|T|-1)!}{|F|!}
  \bigl[\hat v(T\cup\{i\})-\hat v(T)\bigr].
  \label{eq:shapley}
\end{equation}
The Shapley value serves as a marginal-contribution benchmark for total farmer surplus. It satisfies efficiency, symmetry, the dummy axiom, and additivity, but it is not guaranteed to satisfy \eqref{eq:stab}. We therefore check core membership after computing it, using \eqref{eq:eff}--\eqref{eq:stab}.

\paragraph{Attributable-revenue split.}
In this baseline, each farmer receives the revenue generated on her own land and pays her area-proportional certification outlay:
\begin{equation}
  x^{\mathrm{AR}}_i=R_i(\Pi^F_i)-\kappa_i(F),\qquad\forall i\in F.
  \label{eq:ar}
\end{equation}
This is efficient, since $\sum_i x^{\mathrm{AR}}_i=R(F,\Pi^F)-C(F)=p(F)$. It is consistent with the self-reliance screen, but it does not account for coalitional stability.

\paragraph{Area-proportional split.}
In this baseline, the pooled surplus is divided in proportion to farm area:
\begin{equation}
  x^{\mathrm{PS}}_i=\frac{\mathrm{FS}_i}{A(F)}\,p(F),\qquad\forall i\in F.
  \label{eq:ps}
\end{equation}
It is a simple land-based benchmark that ignores differences in carbon mitigation potential across farms.

\subsection{Modeling Assumptions and Limitations}
\label{sec:limits}
\begin{itemize}
  \item \textbf{Coalition-selected portfolios.} The coalition selects the profile that maximizes total surplus, which need not coincide with the portfolio each member would pick if choosing only for herself. 
  \item \textbf{Implementation.} $x_i$ does not depend on realized practices, so we assume members implement the selected JPO portfolio, enforced by contract or by result-based verification through MRV.
  \item \textbf{Single period.} Costs and payments are modeled for one season, without recurring certification or payment lags.
  \item \textbf{Farmer heterogeneity.} Private economics differ across farmers through area, portfolio, and $\mathrm{CMP}_{ij}$. Currently $\mathrm{OC}_{ij}$ and $Y_{ij}$ are not considered but the model can support it.
\end{itemize}

\section{Superadditivity and Convexity}
\label{sec:structure}

In this section, we study two structural properties of the total-surplus game $(F,v)$. We show that the game is superadditive on feasible coalitions under an explicit affordability condition, and that it is not convex in general.

\subsection{Notation and Lemmas}
For a coalition $S$, we denote its total area, total certification cost, and average certification cost per hectare by
\begin{align*}
  A(S) &= \sum_{i\in S}\mathrm{FS}_i,\\
  C(S) &= \mathrm{Fix}_{\mathrm{MRV}}+\mathrm{Var}_{\mathrm{MRV}}A(S)^{\delta}
          +\mathrm{Fix}_T+\mathrm{Var}_T|S|,\\
  \bar c(S) &= \frac{C(S)}{A(S)}.
\end{align*}
For a farmer $i$ with portfolio $\Pi_i$, we define two per-hectare quantities:
\begin{align}
  \theta_i(\Pi_i) &= \frac{B_i}{\mathrm{FS}_i}+\Delta Y(\Pi_i)-\mathrm{OC}(\Pi_i)
      &&\text{(net funds per hectare)},\\
  \eta_i(\Pi_i) &= \mathrm{CMP}_i(\Pi_i)\,\mathrm{CCP}+\Delta Y(\Pi_i)-\mathrm{OC}(\Pi_i)
      &&\text{(gross economic benefit per hectare)}.
\end{align}
Here $\eta_i(\Pi_i)=\bigl(R_i(\Pi_i)+g_i(\Pi_i)\bigr)/\mathrm{FS}_i$. The \emph{affordability margin} of farmer $i$ is
\begin{equation}
  m_i(\Pi_i)=\min\{\theta_i(\Pi_i),\,\eta_i(\Pi_i)\}.
  \label{eq:margin}
\end{equation}
Since $\kappa_i(S)=\mathrm{FS}_i\,\bar c(S)$, dividing the budget constraint \eqref{eq:budget} by $\mathrm{FS}_i$ gives $\theta_i(\Pi_i)\ge\bar c(S)$, and dividing the self-reliance constraint \eqref{eq:selfrel} by $\mathrm{FS}_i$ gives $\eta_i(\Pi_i)\ge\bar c(S)$. Together, the two constraints are equivalent to the per-hectare condition
\begin{equation}
  m_i(\Pi_i)\ \ge\ \bar c(S)\qquad\text{for every } i\in S.
  \label{eq:perha}
\end{equation}
In other words, every member of the coalition must have, both in available funds and in the benefit generated on her own land, at least as much per hectare as the average certification cost per hectare of the coalition.

A coalition $S$ is called \emph{feasible} if some admissible profile satisfies \eqref{eq:perha}. For an infeasible coalition, $\mathrm{OPT}(S)=-\infty$ and, by \eqref{eq:v}, $v(S)=0$, which means that it does not certify. Since $\bar c(S)>0$ and $\eta_i(\emptyset)=0$, the empty portfolio is never admissible for a member of a certifying coalition.

\begin{lemma}[Concavity]
\label{lem:conc}
For $\delta\in(0,1)$ and $a,b>0$,
\[
  (a+b)^{\delta}<a^{\delta}+b^{\delta}.
\]
\end{lemma}
\begin{proof}
Let $\varphi(t)=t^{\delta}$. It is strictly concave because
\[
  \varphi''(t)=\delta(\delta-1)t^{\delta-2}<0,
\]
and $\varphi(0)=0$. Put $\lambda=a/(a+b)$. Concavity gives
\begin{align*}
  \varphi(a) &> \lambda\,\varphi(a+b),\\
  \varphi(b) &> (1-\lambda)\,\varphi(a+b).
\end{align*}
Adding the two inequalities gives the claim.
\end{proof}

\begin{lemma}[Merging reduces total and average certification cost]
\label{lem:merge}
Let $S,T$ be disjoint coalitions, $Q=S\cup T$, $a=A(S)$, $b=A(T)$ and $F_0=\mathrm{Fix}_{\mathrm{MRV}}+\mathrm{Fix}_T$. Then
\[
  \Gamma(S,T):=C(S)+C(T)-C(Q)
  =F_0+\mathrm{Var}_{\mathrm{MRV}}\bigl[a^{\delta}+b^{\delta}-(a+b)^{\delta}\bigr]>0,
\]
and consequently
\begin{equation}
  \bar c(Q)<\frac{a\,\bar c(S)+b\,\bar c(T)}{a+b}\le\max\{\bar c(S),\bar c(T)\}.
  \label{eq:avgcost}
\end{equation}
\end{lemma}
\begin{proof}
\emph{The saving.} The MRV terms contribute
\[
  2\mathrm{Fix}_{\mathrm{MRV}}+\mathrm{Var}_{\mathrm{MRV}}(a^{\delta}+b^{\delta})
  -\mathrm{Fix}_{\mathrm{MRV}}-\mathrm{Var}_{\mathrm{MRV}}(a+b)^{\delta}.
\]
For the transaction terms, $|Q|=|S|+|T|$, so exactly one $\mathrm{Fix}_T$ is saved. Adding the two contributions gives the formula for $\Gamma$.

\emph{Positivity.} By Lemma~\ref{lem:conc}, the bracket is positive. We assume that $\mathrm{Var}_{\mathrm{MRV}}>0$ or $F_0>0$, so $\Gamma>0$.

\emph{The average cost.} Since $\Gamma>0$,
\[
  C(Q)<C(S)+C(T)=a\bar c(S)+b\bar c(T).
\]
Dividing by $a+b$ gives the first inequality in \eqref{eq:avgcost}. The second holds because a weighted average never exceeds the maximum.
\end{proof}

Intuitively, after a merger, the average cost per hectare falls below that of the more expensive side, so that side can always afford the merged coalition. Only the cheaper side may lose affordability, because it now pays a higher average cost than before.

\begin{lemma}[Feasible coalitions have non-negative surplus]
\label{lem:nonneg}
If $S$ is feasible and $\{\Pi_i\}_{i\in S}$ is any admissible profile satisfying \eqref{eq:perha}, then
\[
  \sum_{i\in S}\bigl[R_i(\Pi_i)+g_i(\Pi_i)\bigr]-C(S)\ge 0.
\]
Hence, for every feasible $S$,
\[
  v(S)=\mathrm{OPT}(S)=\sum_{i\in S}\bigl[R_i(\Pi^S_i)+g_i(\Pi^S_i)\bigr]-C(S).
\]
\end{lemma}
\begin{proof}
The self-reliance constraint \eqref{eq:selfrel} gives $R_i(\Pi_i)+g_i(\Pi_i)\ge\kappa_i(S)$ for each $i\in S$. Summing over $i\in S$ and using $\sum_{i\in S}\kappa_i(S)=C(S)$ gives the first claim. The second follows from \eqref{eq:v}.
\end{proof}

\subsection{Superadditivity}
\begin{definition}[Superadditivity]
$(F,v)$ is superadditive on a family of coalitions if $v(S\cup T)\ge v(S)+v(T)$ for all disjoint $S,T$ in the family such that $S\cup T$ also belongs to it.
\end{definition}

\begin{theorem}[Superadditivity]
\label{thm:superadd}
Let $S,T$ be disjoint, non-empty, feasible coalitions with optimal profiles $\Pi^*_S,\Pi^*_T$, labelled so that $\bar c(S)\le\bar c(T)$. Define
\[
  m_S=\min_{i\in S}m_i(\Pi^*_{S,i}),\qquad \Delta_S(T)=C(S\cup T)-C(S).
\]
If
\begin{equation}
  \text{(H)}\qquad \bar c(S\cup T)\le m_S,
  \label{eq:H}
\end{equation}
then the concatenated profile $\tilde\Pi_Q$ is feasible for $Q=S\cup T$, and
\[
  v(S\cup T)\ \ge\ v(S)+v(T)+\Gamma(S,T)\ >\ v(S)+v(T).
\]
Condition (H) holds in particular in either of the following cases:
\begin{enumerate}
  \item $\bar c(S)=\bar c(T)$;
  \item the condition
  \begin{equation}
    \text{(H$'$)}\qquad \Delta_S(T)\le A(T)\,m_S,
    \label{eq:Hprime}
  \end{equation}
  that is, the extra cost of absorbing $T$, per hectare of $T$, is at most the tightest per-hectare affordability margin in $S$.
\end{enumerate}
\end{theorem}

The idea of the proof is simple. We keep every farmer's portfolio unchanged. The merged coalition is cheaper per hectare than its more expensive part, so the farmers of $T$ remain feasible, and (H) is exactly what the farmers of $S$ need.

\begin{proof}
Let every farmer keep her own portfolio, so that every $m_i$ is unchanged. By \eqref{eq:perha}, it suffices to show that $m_i\ge\bar c(Q)$ for all $i\in Q$.

\medskip
\noindent\emph{Step 1: farmers of $T$.}
Since $\Pi^*_T$ is feasible for $T$, we have $m_i\ge\bar c(T)$. By \eqref{eq:avgcost} and the labelling,
\[
  \bar c(T)=\max\{\bar c(S),\bar c(T)\}>\bar c(Q).
\]
Hence $m_i>\bar c(Q)$.

\medskip
\noindent\emph{Step 2: farmers of $S$.}
For $i\in S$, the definition of $m_S$ and condition (H) give
\[
  m_i\ge m_S\ge\bar c(Q).
\]

\medskip
\noindent\emph{Step 3: value.}
The per-farmer benefit $R_i(\Pi_i)+g_i(\Pi_i)$ is additive across farmers and depends only on farmer $i$'s own portfolio, which is unchanged in $\tilde\Pi_Q$. The certification cost satisfies $C(Q)=C(S)+C(T)-\Gamma$. By Lemma~\ref{lem:nonneg}, $v(S)=\sum_{i\in S}\bigl[R_i(\Pi^*_{S,i})+g_i(\Pi^*_{S,i})\bigr]-C(S)$, and likewise for $T$. Therefore
\begin{align*}
  v(Q) &\ge \sum_{i\in Q}\bigl[R_i(\tilde\Pi_{Q,i})+g_i(\tilde\Pi_{Q,i})\bigr]-C(Q)\\
       &= v(S)+v(T)+\Gamma.
\end{align*}

\medskip
\noindent\emph{Special case (a): $\bar c(S)=\bar c(T)$.}
Then \eqref{eq:avgcost} gives $\bar c(Q)<\bar c(S)$. Feasibility of $S$ gives $\bar c(S)\le m_S$, so $\bar c(Q)<m_S$, which is (H).

\medskip
\noindent\emph{Special case (b): (H$'$) implies (H).}
Feasibility of $S$ gives $m_S\ge\bar c(S)$, hence $C(S)\le A(S)\,m_S$. Then
\begin{align*}
  C(Q) &= C(S)+\Delta_S(T)\\
       &\le A(S)\,m_S+A(T)\,m_S\\
       &= A(Q)\,m_S,
\end{align*}
and dividing by $A(Q)$ gives (H).
\end{proof}

\begin{remark}[Why a condition is needed]
\label{rem:cond}
Without (H), the conclusion can fail even when $Q$ is feasible. We illustrate this with a small instance. Take $\delta=\tfrac12$, $\mathrm{Fix}_{\mathrm{MRV}}=\mathrm{Fix}_T=0$, $\mathrm{Var}_{\mathrm{MRV}}=\mathrm{Var}_T=10$, and no yield effect. There are two mutually exclusive practices: $P_2$ with $\mathrm{OC}=0$ and $P_1$ with $\mathrm{OC}=0.5$ \$/ha.
\begin{itemize}
  \item $S=\{F_0\}$ with $\mathrm{FS}=100$ ha and $B/\mathrm{FS}=2.5$. Carbon revenue is $3$ \$/ha under $P_2$ and $10$ \$/ha under $P_1$.
  \item $T$ consists of ten farmers with $\mathrm{FS}=0.1$ ha each, $B/\mathrm{FS}=200$, and carbon revenue of $1000$ \$/ha under either practice.
\end{itemize}

\begin{center}
\renewcommand{\arraystretch}{1.2}
\begin{tabular}{|l|r|r|r|p{6.2cm}|}
\hline
coalition & $A$ & $C$ & $\bar c$ & outcome\\
\hline
$S$ & 100 & 110 & 1.1 & $F_0$ uses $P_1$ ($\theta=2.0\ge 1.1$); $v=950-110=840$\\
$T$ & 1 & 110 & 110 & all use $P_2$ (same revenue, lower cost); $v=1000-110=890$\\
$Q=S\cup T$ & 101 & $\approx 210.50$ & $\approx 2.084$ & $P_1$ unaffordable for $F_0$ ($2.0<2.084$), so $F_0$ uses $P_2$\\
\hline
\end{tabular}
\end{center}
In $Q$, the total benefit $\sum_{i\in Q}\bigl[R_i+g_i\bigr]$ is $300+1000=1300$, so
\[
  v(Q)=1300-210.50=1089.50\ <\ 1730=v(S)+v(T).
\]
The reason is that the many tiny farms raise the average cost per hectare and push the large farmer below the affordability threshold. Here $m_S=\min\{2.0,\,9.5\}=2.0<\bar c(Q)$, so (H) fails.

\end{remark}

\subsection{Convexity}
\begin{definition}[Convexity]
$(F,v)$ is convex if for all $S\subseteq T\subseteq F$ and $i\notin T$,
\[
  v(T\cup\{i\})-v(T)\ \ge\ v(S\cup\{i\})-v(S).
\]
\end{definition}
Convexity says that a farmer's marginal contribution never decreases as the coalition grows. Writing $\Delta_i(X)=v(X\cup\{i\})-v(X)$, adding a farmer $i$ to a coalition has two effects on $\Delta_i$:
\begin{itemize}
  \item a cost saving, which grows with the coalition because of the concave MRV term;
  \item a benefit effect: a larger coalition lowers $\bar c$ and can make costly, high-sequestration practices affordable for other members.
\end{itemize}
The benefit effect can shrink as the coalition grows. For example, a poor farmer who cannot afford a costly practice alone may be unlocked by either of two neighbors, but once one neighbor has joined, the second neighbor unlocks nothing new. This is the mechanism behind the following result.

\begin{theorem}[The game is not convex in general]
\label{thm:nonconvex}
For $\delta=\tfrac12$ there is an instance within the model in which $(F,v)$ is not convex.
\end{theorem}

\begin{proof}
\emph{Instance.} Let $\mathrm{Fix}_{\mathrm{MRV}}=\mathrm{Fix}_T=0$, $\mathrm{Var}_{\mathrm{MRV}}=\mathrm{Var}_T=10$, $\delta=\tfrac12$ and $\Delta Y\equiv 0$. There are two mutually exclusive practices:
\begin{itemize}
  \item $P_2$ earns $40$ \$/ha in carbon revenue with $\mathrm{OC}=0$, so its net benefit is $40$ \$/ha;
  \item $P_1$ earns $100$ \$/ha with $\mathrm{OC}=10$ \$/ha, so its net benefit is $90$ \$/ha.
\end{itemize}
The farmers $(\mathrm{FS}_i,B_i)$ are
\[
  F_1=(1,25),\qquad F_2=(24,1920),\qquad F_3=(24,480).
\]
By \eqref{eq:perha}, farmer $i$ may use practice $p$ in coalition $X$ if and only if
\[
  \frac{B_i}{\mathrm{FS}_i}-\mathrm{OC}_p\ge\bar c(X)
  \quad\text{and}\quad
  \eta_p\ge\bar c(X),
  \qquad\text{where } \eta_{P_1}=90,\ \eta_{P_2}=40.
\]
We take
\[
  S=\{F_1\}\ \subset\ T=\{F_1,F_2\},\qquad i=F_3\notin T.
\]

\medskip
\emph{Values.} Here
\[
  C(X)=10\sqrt{A(X)}+10|X|.
\]
Since $P_1$ has the higher net benefit per hectare, the optimal profile is all-$P_1$ whenever that is feasible.

\begin{center}
\renewcommand{\arraystretch}{1.2}
\begin{tabular}{|l|r|r|r|p{5.2cm}|r|}
\hline
coalition & $A$ & $C$ & $\bar c$ & optimal profile & $v$\\
\hline
$S=\{F_1\}$ & 1 & 20 & 20 & $F_1$ uses $P_2$ ($P_1$: $\theta=25-10=15<20$) & $40-20=20$\\
$S\cup\{F_3\}$ & 25 & 70 & 2.8 & all $P_1$ ($\theta=15,10\ge 2.8$) & $2250-70=2180$\\
$T=\{F_1,F_2\}$ & 25 & 70 & 2.8 & all $P_1$ ($\theta=15,70\ge 2.8$) & $2250-70=2180$\\
$T\cup\{F_3\}$ & 49 & 100 & $\approx 2.04$ & all $P_1$ & $4410-100=4310$\\
\hline
\end{tabular}
\end{center}
In the last three rows, all-$P_1$ is feasible and has the largest net benefit, so it is optimal. In the first row, $P_1$ is infeasible for $F_1$, so $F_1$ must use $P_2$.

\medskip
\emph{Violation.}
\begin{align*}
  \Delta_3(S) &= 2180-20 = 2160,\\
  \Delta_3(T) &= 4310-2180 = 2130.
\end{align*}
Since $S\subseteq T$ and $F_3\notin T$ but $\Delta_3(T)<\Delta_3(S)$, convexity fails.
\end{proof}


\section{Experimental Results}
\label{sec:results}
This section evaluates the proposed framework through controlled computational experiments. We first describe \textsc{CarbonFARM-Bench}, the expert-constructed dataset on which all experiments are run, and then the experimental setup. We then present results on the structural barriers to solo participation; the effects of certification costs, farm scale, and carbon prices on coalition viability; the stability of alternative allocation mechanisms; and the resulting policy implications.

\subsection{Dataset}
\label{sec:dataset}

All experiments in this paper use \textsc{CarbonFARM-Bench} (\textbf{F}air \textbf{A}llocation and \textbf{R}evenue \textbf{M}aximization for Smallholders in Carbon Farming Benchmark), an expert-constructed carbon farming dataset for smallholder systems, which we release alongside this work.\footnote{The dataset is publicly available at \url{https://github.com/AIMLGT-Agriculture-2/CarbonFARM-Bench}.} \textsc{CarbonFARM-Bench} was assembled by domain experts who consolidated agronomic, economic, and carbon-market evidence from the literature into a single, internally consistent benchmark for studying coalition formation among smallholder farmers. Because published sources report these quantities in heterogeneous units, regions, and contexts, expert curation was needed to reconcile them into one coherent parameter set. The full construction procedure, including parameter tables, interaction matrices, sampling distributions, and farmer panels, is described in Appendix~\ref{appendix:dataset}. The main components are summarized below.

\begin{itemize}
    \item \textbf{Practice-level data.} The dataset covers $m=8$ rice-specific and common carbon sequestration practices. Each practice is characterized by its sequestration potential, operational cost, and yield impact, with values grounded in \cite{das2025enhancing,tyagi2025climate,agronomy14020378}.

    \item \textbf{Pairwise interaction matrices.} Three symmetric $8\times 8$ matrices ($\alpha$, $\beta$, $\gamma$) encode the effects of co-adopting practices on sequestration, cost, and yield, respectively. They capture agronomic synergies and antagonisms that cannot be inferred from practice-level values alone.

    \item \textbf{Farmer-level data.} The base panel contains $n=15$ smallholder farmers. Farm sizes are drawn from a truncated log-normal distribution over $0.10$--$2.00$\,ha, following observed smallholder farm-size distributions \citep{lowder2016number}, and budgets are positively correlated with farm size. We use $n=15$ because it brings out the conceptual findings most clearly. To show that these findings scale, we also report results on a larger panel of $n=250$ farmers.

    \item \textbf{Economic and MRV parameters.} Carbon credit prices and the costs of monitoring, reporting, and verification (MRV) and of transactions are set to the baseline values in Table~\ref{tab:config_params}.
\end{itemize}

\begin{table}[ht]
\centering
\caption{Baseline configuration parameters in \textsc{CarbonFARM-Bench}.}
\vspace{1em}
\small
\label{tab:config_params}
\setlength{\tabcolsep}{4pt}
\begin{tabular}{llrl}
\hline
\textbf{Symbol} & \textbf{Description} & \textbf{Baseline} & \textbf{Unit} \\
\hline
$\text{CCP}$ & Carbon credit price & $60$ & \$/tCO$_2$e \\
$\text{Fix}_{\text{MRV}}$ & Fixed cost of MRV & $2000$ & \$ \\
$\text{Var}_{\text{MRV}}$ & Variable cost of MRV & $80$ & \$/ha$^{\delta}$ \\
$\delta$ & MRV scale exponent & $0.70$ & --- \\
$\text{Fix}_{\text{T}}$ & Fixed transaction cost & $1000$ & \$ \\
$\text{Var}_{\text{T}}$ & Variable transaction cost & $10$ & \$/farmer \\
$\bar{b}$ & Mean farmer budget & $140$ & \$/ha \\
\hline
\end{tabular}
\end{table}
 
\subsection{Experimental Setup}
All experiments use $n = 15$ farmers (with the base configuration parameters in Table~\ref{tab:config_params}) and are averaged over 10 independently sampled farmer panels. We use $n=15$ as an exact-computation benchmark because exhaustive evaluation of all $2^{n}-1$ non-empty coalitions is tractable at this scale. This enables exact computation and verification of core-related properties and of all solution concepts.

The numerical experiments are intended as controlled computational experiments for evaluating the structural properties of the proposed coalition model, and for obtaining empirical estimates of carbon-farming profitability under the stated assumptions.

We report the following quantities throughout, The \emph{solo value} $\sum_i v(\{i\})$ is the non-cooperative benchmark which tells whether farmers can participate individually. The grand-coalition value $v(N)$ is the total surplus created when everyone certifies together. It splits into two parts: one, the pooled component $p(N)$, which is carbon revenue net of certification costs and is the \emph{transferable} surplus that the allocation rules divide among members, and second, the private component $G(N)$, which collects the agronomic gains (yield and input-cost effects) that each farmer keeps irrespective of the allocation. 

The cooperative gain $\Delta v = v(N) - \sum_i v(\{i\})$ measures what is created by cooperating over and above what farmers can obtain alone, and its relative size $\Delta v / v(N)$ shows how much of the surplus is due to coalition formation. Finally, since the grand coalition may itself fail to meet the certification and affordability constraints for some parameter values, we report the fraction of panels on which it certifies, so that $v(N)$ and $p(N)$ are always read together with the probability that the coalition exists.

\subsection{Structural Barriers to Solo Participation}
\label{sec:solo}

Table~\ref{tab:budget} varies the mean budget per hectare $\bar b$ from $0$ to $1000$ \$/ha, and Table~\ref{tab:scale} multiplies all farm areas by a factor $k$ while holding the budget per hectare fixed.

\begin{table}[ht]
\centering
\caption{Effect of the mean budget $\bar b$, $^\ast$ represents the base configuration parameter value.}
\label{tab:budget}
\begin{tabular}{rrrrrr}
\toprule
$\bar b$ (\$/ha) & Solo-feasible & $\sum_i v(\{i\})$ & $v(N)$ & $p(N)$ & Grand feasible \\
\midrule
0    & 0 & 0 & 5{,}600 & 1{,}356 & 0.90 \\
25   & 0 & 0 & 5{,}600 & 1{,}356 & 0.90 \\
50   & 0 & 0 & 5{,}918 & 1{,}345 & 1.00 \\
75   & 0 & 0 & 5{,}929 & 1{,}358 & 1.00 \\
140$^\ast$ & 0 & 0 & 5{,}929 & 1{,}358 & 1.00 \\
400  & 0 & 0 & 5{,}929 & 1{,}358 & 1.00 \\
1000 & 0 & 0 & 5{,}929 & 1{,}358 & 1.00 \\
\bottomrule
\end{tabular}
\end{table}

\begin{table}[ht]
\centering
\caption{Effect of farm scale $k$ (areas multiplied by $k$), $^\ast$ represents the base configuration parameter value.}
\label{tab:scale}
\begin{tabular}{rrrrrrr}
\toprule
$k$ & Avg.\ size (ha) & Solo-feasible & $\sum_i v(\{i\})$ & $v(N)$ & $\Delta v$ & Grand feasible \\
\midrule
0.5  & 0.34  & 0    & 0       & 471     & 471     & 0.20 \\
1$^\ast$ & 0.67 & 0 & 0       & 5{,}929 & 5{,}929 & 1.00 \\
2    & 1.34  & 0    & 0       & 15{,}159 & 15{,}159 & 1.00 \\
3    & 2.01  & 0.2  & 542     & 24{,}426 & 23{,}883 & 1.00 \\
5    & 3.35  & 3.0  & 9{,}389 & 43{,}014 & 33{,}626 & 1.00 \\
10   & 6.70  & 8.1  & 46{,}353 & 89{,}643 & 43{,}290 & 1.00 \\
20   & 13.40 & 12.4 & 135{,}822 & 183{,}191 & 47{,}369 & 1.00 \\
\bottomrule
\end{tabular}
\end{table}

\subsubsection*{Insights}
\paragraph{Budget.} No farmer can certify alone at any budget level, even at more than seven times the baseline. At the baseline certification cost, a farm of under two hectares cannot generate enough carbon credits to satisfy the self-reliance constraint (\eqref{eq:selfrel}), making this constraint the primary reason that rules out solo participation. Whereas the coalition is insensitive to the budget once $\bar b \ge 75$ \$/ha. The budget constraint (\eqref{eq:budget}) binds on the grand coalition only at very small budgets; when $\bar b \le 25$, it fails to certify in $10\%$ of panels.

\paragraph{Farm scale.} Solo feasibility emerges only once the average farm size grows severalfold. The grand coalition creates more value than the sum of solo values at every scale, but the \emph{relative} gain($\Delta v / v(N)$) shrinks: It falls from $100\%$ for $k \le 2$ to $26\%$ at $k=20$. Conversely, halving farm sizes makes even the grand coalition unviable in $80\%$ of panels. Coalition formation is thus most valuable where farms are smallest, yet its gain persists even for large farms.

\medskip
\noindent Together, Tables~\ref{tab:budget} and~\ref{tab:scale} point to a \emph{structural} barrier: budget increases do not relax the binding constraint on solo farmers, farm size does. This gives a clear economic rationale for coalition formation as an institutional necessity for smallholders.

\subsection{Effect of Certification Costs}
\label{sec:costs}

\begin{table}[ht]
\centering
\caption{Effect of certification-cost parameters, $^\ast$ represents the base configuration parameter value.}
\label{tab:costs}
\begin{tabular}{llrrr}
\toprule
Parameter & Value & $v(N)$ & $p(N)$ & Grand feasible \\
\midrule
\multirow{6}{*}{$\FixMRV$ (\$)}
 & 0       & 7{,}929 & 3{,}358 & 1.00 \\
 & 1000    & 6{,}929 & 2{,}358 & 1.00 \\
 & 2000$^\ast$ & 5{,}929 & 1{,}358 & 1.00 \\
 & 3000    & 4{,}086 & 573     & 0.70 \\
 & 4000    & 2{,}321 & 112     & 0.40 \\
 & 6000    & 0       & 0       & 0.00 \\
\midrule
\multirow{5}{*}{$\FixT$ (\$)}
 & 0       & 6{,}929 & 2{,}358 & 1.00 \\
 & 1000$^\ast$ & 5{,}929 & 1{,}358 & 1.00 \\
 & 2000    & 4{,}086 & 573     & 0.70 \\
 & 4000    & 1{,}580 & $-144$  & 0.30 \\
 & 6000    & 0       & 0       & 0.00 \\
\midrule
\multirow{4}{*}{$\VarMRV$ (\$/ha$^{\delta}$)}
 & 0       & 6{,}330 & 1{,}759 & 1.00 \\
 & 80$^\ast$ & 5{,}929 & 1{,}358 & 1.00 \\
 & 160     & 5{,}519 & 946     & 1.00 \\
 & 240     & 4{,}844 & 598     & 0.90 \\
\midrule
\multirow{6}{*}{$\VarT$ (\$/farmer)}
 & 0       & 6{,}079 & 1{,}508 & 1.00 \\
 & 10$^\ast$ & 5{,}929 & 1{,}358 & 1.00 \\
 & 40      & 5{,}195 & 951     & 0.90 \\
 & 80      & 4{,}051 & 538     & 0.70 \\
 & 160     & 2{,}199 & $-13$   & 0.40 \\
 & 320     & 0       & 0       & 0.00 \\
\bottomrule
\end{tabular}
\end{table}

\subsubsection*{Insights}

\paragraph{Fixed costs.} Around the baseline, $v(N)$ moves one-for-one with both $\FixMRV$ and $\FixT$ (each \$1000 change in either shifts $v(N)$ and $p(N)$ by \$1000). Above the baseline the effect becomes non-linear because the \emph{affordability} constraints start to bind and also effects the grand coalition feasibility.

\paragraph{Variable MRV.} Tripling $\VarMRV$ from $80$ to $240$ lowers $v(N)$ by $18\%$ and begins to threaten feasibility. Whereas setting $\VarMRV = 0$ raises $v(N)$ by $7\%$.

\paragraph{Variable transaction cost.} Around the baseline, each extra dollar of $\VarT$ lowers $v(N)$ by \$15, one dollar for each of the $15$ members. Because this cost is paid per farmer, it does not dilute as the coalition grows, and feasibility starts to erode at $\VarT \ge 40$.

\noindent  Together we observe that fixed costs and the variable transaction cost $\VarT$ drive coalition viability, while $\VarMRV$ is the mildest lever. Subsidizing fixed and per-farmer transaction costs is the most effective way to keep certification viable for smallholder coalitions.

\subsection{Effect of the MRV Scale Exponent}
\label{sec:delta}

\begin{table}[ht]
\centering
\caption{Effect of the MRV scale exponent $\delta$, $^\ast$ represents the base configuration parameter value.}
\label{tab:delta}
\begin{tabular}{rrrr}
\toprule
$\delta$ & $v(N)$ & $p(N)$  \\
\midrule
0.0 & 6{,}250 & 1{,}679 \\
0.3 & 6{,}171 & 1{,}600  \\
0.5 & 6{,}078 & 1{,}506  \\
0.7$^\ast$ & 5{,}929 & 1{,}358 \\
0.9 & 5{,}693 & 1{,}121  \\
1.0 & 5{,}526 & 955    \\
\bottomrule
\end{tabular}
\end{table}

\subsubsection*{Insights}
As $\delta$ approaches $1$, the economies of scale in monitoring disappear, and the value of the grand coalition decreases monotonically.  
Even at $\delta=1$, the grand coalition retains most of its value, since pooling still removes duplicated fixed costs. The effect is much larger on the \emph{transferable} surplus $p(N)$ that is actually divided among farmers, which falls by $43\%$ over the same range. Preserving sub-linear MRV costs therefore matters most when the pooled surplus is thin. This makes a case for public support for MRV technologies and institutions, such as shared remote-sensing infrastructure, that keep monitoring costs sublinear.

\subsection{Effect of the Number of Farmers}
\label{sec:population}

\begin{table}[ht]
\centering
\caption{Effect of Coalition  Size.}
\label{tab:population}
\begin{tabular}{rrrrrr}
\toprule
$n$ & $v(N)$ & $v(N)/n$ & $p(N)/n$ & Cert.\ cost/ha & Solo-feasible \\
\midrule
5    & 0        & ---   & ---   & ---    & 0 \\
10   & 2{,}464  & 246   & 27    & 432.2  & 0 \\
15   & 5{,}929  & 395   & 91    & 365.8  & 0 \\
25   & 11{,}775 & 471   & 168   & 236.5  & 0 \\
50   & 27{,}702 & 554   & 243   & 131.5  & 0 \\
100  & 58{,}533 & 585   & 277   & 81.9   & 0 \\
250  & 149{,}095 & 596  & 294   & 50.3   & 0 \\
500  & 307{,}758 & 616  & 308   & 37.6   & 0 \\
1000 & 615{,}553 & 616  & 310   & 30.7   & 0 \\
\bottomrule
\end{tabular}
\end{table}

\subsubsection*{Insights}
As the number of farmers increases the certification cost per hectare decreases (\$432 at $n=10$ to \$31 at $n=1000$), where as the total and the per farmer surplus rise and then flatten: $v(N)/n$ goes from \$246 ($n=10$) to \$554 ($n=50$), \$596 ($n=250$) and \$616 ($n=1000$). Most of the scale benefit is thus realised with a few dozen to a few hundred members, which is consistent with the size of typical farmer producer organisations.

\subsection{Effect of the Carbon Credit Price}
\label{sec:ccp}

\begin{table}[ht]
\centering
\caption{Effect of the carbon credit price, $^\ast$ represents the base configuration parameter value.}
\label{tab:ccp}
\begin{tabular}{crr}
\toprule
$\CCP$ (\$/tCO$_2$e) & $v(N)$ & $p(N)$ \\
\midrule
10  & 1{,}842  & $-2{,}750$ \\
30  & 3{,}475  & $-1{,}096$ \\
50  & 5{,}111  & 540        \\
60$^\ast$ & 5{,}929 & 1{,}358  \\
100 & 9{,}202  & 4{,}631   \\
150 & 13{,}293 & 8{,}721    \\
200 & 17{,}383 & 12{,}812  \\
300 & 25{,}565 & 20{,}993  \\
500 & 41{,}927 & 37{,}356   \\
\bottomrule
\end{tabular}
\end{table}

\subsubsection*{Insights}
The total surplus $v(N)$ rises roughly linearly with the price. The pooled surplus $p(N)$ becomes positive only above about \$43/tCO$_2$e. Below that price $p(N) < 0$: carbon revenue does not cover certification, and the coalition is kept feasible by the farmers' \emph{private} agronomic gains $G(N)$. The feasibility of the grand coalition at low prices should therefore be read with care as it can be driven by yield and cost co-benefits of the practices as much as by carbon revenue. A programme that relies on carbon income alone would need a price of roughly \$43 or more at baseline costs.

\subsection{Comparison of Revenue Allocation Mechanisms}
\label{sec:exp-allocation}

We compare six allocation mechanisms: the Shapley value, the \emph{core} (represented by its centroid, estimated by hit-and-run sampling), the least core, the nucleolus, and the two baselines, namely the area-proportional split and the attributable-revenue split. Allocations are reported as final surplus $u_i = x_i + g_i(\Pi_i^N)$. Table~\ref{tab:stability} reports stability at the baseline, table reports ~\ref{tab:payoffs} the final payoffs for one panel and its equity indicators.

\begin{table}[ht]
\centering
\caption{Stability at the baseline. Max excess is $\max_T [\hat v(T) - x(T)]$ (negative implies stablility); min gain is the smallest improvement over solo.}
\label{tab:stability}
\begin{tabular}{lrrrr}
\toprule
Mechanism & Stable (frac.) & Violations & Mean max excess (\$) & Mean min gain (\$) \\
\midrule
Shapley             & 1.00 & 0 & $-53.6$ & 85.5 \\
Core (centroid)     & 1.00 & 0 & $-70.2$ & 78.6 \\
Least core          & 1.00 & 0 & $-74.6$ & 74.6 \\
Nucleolus           & 1.00 & 0 & $-74.6$ & 74.6 \\
Area-proportional   & 1.00 & 0 & $-43.9$ & 101.8 \\
Attributable-revenue & 1.00 & 0 & $-47.1$ & 100.5 \\
\bottomrule
\end{tabular}
\end{table}

\begin{table}[ht]
\centering
\caption{Final payoffs $u_i = x_i + g_i$ (\$) for the panel with seed 42, sorted by farm size. The last row gives the Gini coefficient of $u$ under each mechanism for this panel.}
\label{tab:payoffs}
\small
\begin{tabular}{lrrrrrrrr}
\toprule
Farmer & Size (ha) & Avg CMP & Shapley & Core & Least core & Nucleolus & Area-prop. & Attrib.\ rev. \\
\midrule
F1  & 0.16 & 1.77 & 86.8  & 66.9  & 61.5  & 61.5  & 101.1 & 91.0 \\
F2  & 0.21 & 1.78 & 130.4 & 101.4 & 128.8 & 95.2  & 132.7 & 145.2 \\
F3  & 0.26 & 1.88 & 165.2 & 134.1 & 179.7 & 120.6 & 164.3 & 182.9 \\
F4  & 0.26 & 2.03 & 156.4 & 130.9 & 177.9 & 119.7 & 164.3 & 181.2 \\
F5  & 0.44 & 1.90 & 272.7 & 229.5 & 328.1 & 194.8 & 278.0 & 284.2 \\
F6  & 0.60 & 1.97 & 391.4 & 393.1 & 467.8 & 329.2 & 379.1 & 436.3 \\
F7  & 0.65 & 2.00 & 415.7 & 380.5 & 490.5 & 339.1 & 410.7 & 433.1 \\
F8  & 0.69 & 1.76 & 424.4 & 361.5 & 508.7 & 340.0 & 436.0 & 423.6 \\
F9  & 0.82 & 1.71 & 522.1 & 488.8 & 567.9 & 482.0 & 518.1 & 531.5 \\
F10 & 0.97 & 1.93 & 621.3 & 607.2 & 636.1 & 617.7 & 612.9 & 627.8 \\
F11 & 1.28 & 1.73 & 814.0 & 844.2 & 777.1 & 873.6 & 808.8 & 802.7 \\
F12 & 1.32 & 1.95 & 840.6 & 861.4 & 795.2 & 905.9 & 834.0 & 824.5 \\
F13 & 1.46 & 1.71 & 948.4 & 993.1 & 858.9 & 1{,}055.6 & 922.5 & 937.5 \\
F14 & 1.55 & 1.68 & 983.5 & 1{,}100.8 & 899.8 & 1{,}112.2 & 979.4 & 970.7 \\
F15 & 1.87 & 1.78 & 1{,}150.8 & 1{,}230.1 & 1{,}045.4 & 1{,}276.4 & 1{,}181.6 & 1{,}051.3 \\
\midrule
\multicolumn{3}{l}{Gini$(u)$} & 0.347 & 0.385 & 0.309 & 0.407 & 0.326 & 0.333 \\
\bottomrule
\end{tabular}
\end{table}

\begin{table}[ht]
\centering
\caption{Superadditivity and convexity checked over all coalitions ($n=15$).}
\label{tab:structure}
\begin{tabular}{rrrrr}
\toprule
Panel seed & $v(N)$ & Feasible coalitions & Superadd.\ violations & Convexity violations \\
\midrule
42 & 7{,}923 & 17{,}610 / 32{,}767 & 0  & 13.65\% (234{,}854) \\
43 & 6{,}221 & 6{,}559 / 32{,}767  & 0        & 6.89\% (118{,}484) \\
44 & 8{,}577 & 19{,}824 / 32{,}767 & 0  & 15.55\% (267{,}506) \\
45 & 5{,}377 & 4{,}760 / 32{,}767  & 0      & 5.51\% (94{,}778) \\
46 & 3{,}821 & 1{,}216 / 32{,}767  & 0         & 1.28\% (22{,}028) \\
47 & 4{,}318 & 779 / 32{,}767      & 0        & 0.80\% (13{,}766) \\
48 & 8{,}394 & 17{,}278 / 32{,}767 & 0   & 13.46\% (231{,}576) \\
49 & 6{,}319 & 11{,}341 / 32{,}767 & 0       & 11.14\% (191{,}568) \\
50 & 5{,}052 & 4{,}075 / 32{,}767  & 0       & 4.24\% (72{,}984) \\
51 & 3{,}291 & 87 / 32{,}767       & 0       & 0.05\% (916) \\
\bottomrule
\end{tabular}
\end{table}

\newpage
\subsubsection*{Insights}
\paragraph{Stability.} On all $10$ baseline panels the core is non-empty and in fact large, the least-core slack is $\varepsilon^\ast = -74.6$ on average. Consequently \emph{all six} allocation rules are stable on every panel, with no blocking coalition and no individual-rationality violation. Also on average the worst-off farmer gains are \$74.6 over her solo outcome. The size of the core is consistent with strong economies of scale: most proper subgroups cannot certify at all, so no subgroup can come close to what the grand coalition delivers.

\paragraph{Allocation.} The least core is the most egalitarian of the core-based rules and shifts surplus towards small farms relative to the nucleolus and the core centroid. Its Gini coefficient of final surplus is $0.31$, against $0.41$ for the nucleolus and $0.39$ for the core centroid. The nucleolus and the core centroid concentrate surplus on the largest farms.

\paragraph{Superadditivity and Convexity.} We observe no violation of superadditivity is found in the panels of study. Whereas non-convexity is found across all the panels. The violation rate rises with the number of feasible coalitions (from $916$ violations when only $87$ coalitions are feasible to over $267{,}000$ when $19{,}824$ are), so non-convexity is most pronounced in panels where the certification threshold is crossed by many coalitions.

\subsection{When Do Allocation Rules Become Unstable?}
\label{sec:instability}

At the baseline, all six allocation rules are stable (Table~\ref{tab:stability}). This is a property of the baseline parameters as no coalition can credibly threaten to leave. We therefore ask under what conditions stability breaks down. By Proposition~\ref{prop:block}, an allocation $x$ is blocked by a group $T\subsetneq F$ if and only if $\hat v(T)>\sum_{i\in T}x_i$, so we count the coalitions that satisfy this inequality under each rule.

We vary two parameters that govern how strongly a subgroup can outperform its share of the grand coalition. The first is the carbon credit price $\CCP$, which scales the pooled revenue $R(S,\Pi^S)$ relative to the certification cost $C(S)$. The second is the farmer-specific heterogeneity $\nu$ in carbon mitigation potential, where
\[
  \mathrm{CMP}_{ij}=\mathrm{CMP}_j\,\bigl(1+U(-\nu,\nu)\bigr),
\]
so that a larger $\nu$ means the carbon yield per hectare is more unequal across farms. We compare the area-proportional split and the Shapley value. These are the two rules that do not enforce the stability constraints \eqref{eq:stab} by construction. 

\begin{table}[ht]
\centering
\caption{Instability under CMP heterogeneity $\nu$ and carbon price. ``\# block'' is the mean number of blocking coalitions; ``stable'' is the fraction of panels with no blocking coalition.}
\label{tab:instability}
\begin{tabular}{rr rr rr}
\toprule
 & & \multicolumn{2}{c}{Area-proportional} & \multicolumn{2}{c}{Shapley} \\
\cmidrule(lr){3-4}\cmidrule(lr){5-6}
$\CCP$ & $\nu$ & \# block & stable & \# block & stable \\
\midrule
60  & 0.20 & 0      & 1.00 & 0     & 1.00 \\
60  & 0.60 & 0      & 1.00 & 0     & 1.00 \\
60  & 0.95 & 0.9    & 0.70 & 0     & 1.00 \\
\midrule
150 & 0.20 & 0      & 1.00 & 0     & 1.00 \\
150 & 0.60 & 25.0   & 0.50 & 0     & 1.00 \\
150 & 0.95 & 256.7  & 0.10 & 2.8   & 0.50 \\
\midrule
300 & 0.20 & 1.3    & 0.70 & 0     & 1.00 \\
300 & 0.60 & 524.0  & 0.00 & 14.9  & 0.10 \\
300 & 0.95 & 1{,}278.7 & 0.00 & 184.6 & 0.00 \\
\bottomrule
\end{tabular}
\end{table}

\subsubsection*{Insights}

\paragraph{Area-proportional split.}
The area-proportional rule is the first to fail, and it fails as either heterogeneity or price rises. Since $x^{\mathrm{PS}}_i$ depends only on farm size \eqref{eq:ps}, it pays every hectare the same amount regardless of how much carbon that hectare generates. It thus over-rewards low-CMP land and under-rewards high-CMP land, which gives subgroups with above-average CMP a reason to leave. At $\CCP=60$ the rule is stable on all panels up to $\nu=0.6$. At $\CCP=150$ it is stable on only half of the panels at $\nu=0.6$, and on only $10\%$ at $\nu=0.95$. At $\CCP=300$ it is already unstable on $30\%$ of panels at $\nu=0.2$, and on every panel once $\nu\ge0.6$. Raising the price enlarges the stakes of misallocation, because a larger share of $v(N)$ is pooled carbon revenue.

\paragraph{Shapley value.}
The Shapley value is considerably more robust, because it rewards marginal contribution and therefore partly tracks carbon heterogeneity, but it is not immune. It remains stable on every panel at the baseline price for all $\nu$, and at $\CCP=150$ for $\nu\le0.6$. It loses core membership when high prices coincide with high heterogeneity: at $(\CCP,\nu)=(150,0.95)$ it is stable on $50\%$ of panels, and at $(300,0.6)$ on only $10\%$. This is consistent with the non-convexity result of Theorem~\ref{thm:nonconvex}, since the Shapley value of a non-convex game need not lie in the core.

\paragraph{Structure of blocking coalitions.}
In the panels we inspected, the most violated coalitions are very large, typically $12$--$14$ of the $15$ farmers holding roughly $84$--$96\%$ of the total area, and have a slightly higher mean CMP than the panel average. Instability therefore takes the form of a high-carbon majority leaving one or two low-carbon farmers behind.

These results show that a rule which ignores carbon heterogeneity is fragile exactly where it over-pays the farmers who contribute least. Simple rules are adequate when farms are similar and prices are modest, but as $\nu$ or $\CCP$ rises, allocations must be verified against the stability constraints \eqref{eq:stab}, and core-based rules such as the least core or nucleolus are the safer default.

\subsection{Policy Implications}
\label{sec:policy}

Our findings show that smallholder farmers cannot derive the benefits of carbon farming on their own, because certification (MRV) and transaction costs are too high relative to the credits their farms can generate. Support aimed at individual farmers, such as credit or cash transfers, is therefore unlikely to open participation, whereas aggregation does. Governments could promote farmer collectives to enable group participation. Introducing carbon farming as an additional activity within existing farmer producer organisations may substantially reduce the institutional cost of coalition formation by leveraging structures that already exist, since groups of a few dozen to a few hundred members are already large enough to capture most of the scale benefit. It may also provide a supplementary income stream for member farmers and give governments a ready-made channel for smallholder participation in carbon markets.

Public support is most effective when it targets the costs that decide whether a group can certify at all. Fixed MRV and fixed transaction costs are the main entry barrier, and initial financial support, such as grants, subsidies, and low-interest loans for one-time registry, baseline, and auditing costs, will help smallholder collectives overcome it. Per-farmer administrative costs also deserve attention, as they do not dilute as a coalition grows, and standardised digital enrolment can keep them low. Government support for shared digital MRV systems, through remote-sensing-based tools, automated reporting tools, AI-based techniques, and standardised protocols, can significantly reduce costs and encourage farmers to adopt carbon farming. Keeping monitoring costs sub-linear in farm area matters most when the surplus available to share is thin.

The results also highlight the importance of principled revenue sharing within groups. Simple rules, such as an area-proportional split, are adequate when farms are similar and carbon prices are modest, but they become fragile as carbon heterogeneity or prices rise, and the Shapley value follows. Core-based rules, such as the least core and the nucleolus, remain stable whenever any stable allocation exists and could serve as the default for heterogeneous groups or higher-price settings. Revenue-sharing mechanisms should strike a balance between stability and fairness, ensuring that all farmers have an incentive to remain in the coalition.

\section{Conclusion and Future Work}
This paper presented a cooperative game-theoretic framework for enabling smallholder participation in carbon farming through coalition formation. It links three stages: solo portfolio optimization, joint portfolio optimization, and surplus allocation. The framework respects the fact that farmers pool only certification costs and carbon revenue, and keep their agronomic costs and yields private. Coalition formation offers smallholders a channel to additional income and the sustainability co-benefits of carbon farming.

Our results show that individual participation becomes economically infeasible due to farm size and fixed certification costs, while participating in a coalition of farmers generates non-trivial value through economies of scale. We have demonstrated that coalition value is structurally driven by MRV cost dynamics rather than by available budget levels alone. While land size is a natural and important criterion for allocation, value in carbon farming is created not merely by owning more land but by combining land, technology, infrastructure (MRV), and management.

On the structure of the game, we proved that it is superadditive under an affordability condition on merging coalitions. We also showed by construction that it is non-convex in general. The Shapley value is therefore not guaranteed to lie in the core, and the core itself may be empty in general. We compared the Shapley value, the core centroid, the least core, and the nucleolus against area-proportional and attributable-revenue baselines. At baseline, the core was non-empty and large, and all six rules were stable. Stability broke down as carbon heterogeneity and carbon prices rose. Among the core-based rules, the least core was the most egalitarian, while the nucleolus and core centroid concentrated surplus on larger farms. Stability is thus the property that matters most for sustained participation. The Shapley value is a sound choice when it lies in the core, and the least core or nucleolus is the safer default for heterogeneous groups or higher-price settings.

Overall, the study establishes that carbon farming for smallholders is fundamentally an institutional design problem, requiring aggregation, cost-sharing flexibility, and stability-aware incentive mechanisms for scalable and equitable adoption. Our study yields several policy recommendations for the successful adoption of carbon farming practices by smallholder farmers.


\subsection*{Future Work}
Several promising directions emerge naturally from this work. First, while we have used expert-constructed data for clear reasons, empirical validation using real-world datasets from farmer cooperatives or pilot carbon farming projects would significantly strengthen the model's practical relevance. Second, extending the model to dynamic, multi-season settings would allow the study of trust, learning, and coalition stability over time. Third, incorporating uncertainty in carbon prices, yields, and sequestration could enable analysis of risk-sharing mechanisms and robust coalition design. 

Fourth, our exact computations rely on enumerating all $2^n$ coalitions, which limits the allocation analysis to small panels. Scaling to realistic group sizes calls for sampling-based approximations of the Shapley value, least core, and nucleolus, with probabilistic stability guarantees.  Fifth, we studied only the grand coalition, which is justified under our affordability condition but not unconditionally, since superadditivity can fail when coalitions are infeasible or merging raises some members’ outlays. Examining alternative coalition structures, such as clustered or regional groupings, and cost-sharing rules that guarantee superadditivity would give a fuller picture. Sixth, future work should develop simple, transparent allocation rules that approximate core stability and are easy to deploy in practice. Finally, we assumed that farmers are honest and implement the coalition-selected portfolios, which is enforced by contract or MRV. Relaxing this assumption, including strategic misreporting and portfolio deviations, would motivate mechanism-design approaches to the problem.

\subsection*{Acknowledgments}
The authors thank Prof. Soma Dhawala,   Prof. Lalith Achoth,  Prof. Vedamurthy, Mr. Ravi Trivedi and Dr. Soumya K for their valuable suggestions and agronomic insights.
We have utilized various AI language models and assistants, such as GPT and Claude, to enhance the writing quality and provide summaries of source materials. However, the authors diligently verified all information provided by these AI tools to ensure accuracy and factual integrity.

\bibliography{tmlr}
\bibliographystyle{tmlr}

\newpage
\appendix
\section{Notation Summary}
\label{appendix:notation_summary}

\begin{itemize}
    \item $F=\{1,2,\ldots,n\}$: set of farmers.
    \item $P=\{1,2,\ldots,m\}$: set of carbon sequestration practices.
    \item $F_i$: farmer $i$.
    \item $P_j$: practice $j$.
    \item $S,T \subseteq N$: coalitions of farmers.
    \item $|S|$: size of coalition $S$.

    \item $\mathrm{FS}_i$: farm size of farmer $i$ (ha).
    \item $B_i$: seasonal budget of farmer $i$ (\$ ).
    \item $\mathrm{ha}$: hectare.
    \item $\mathrm{tCO}_2 \mathrm{e}$: tonne of CO$_2$ equivalent.
    \item $\mathrm{CCP}$: carbon credit price (\$/$\mathrm{tCO}_2 \mathrm{e}$).

    \item $\mathrm{OC}_j$: per-hectare operational cost of practice $j$.
    \item $\mathrm{CMP}_j$: baseline additionality-adjusted per-hectare sequestration of practice $j$.
    \item $\mathrm{CMP}_{ij}$: farmer-specific additionality-adjusted per-hectare sequestration of practice $j$ for farmer $i$.
    \item $Y_j$: per-hectare yield revenue effect of practice $j$.

    \item $\Pi \subseteq P$: portfolio of practices.
    \item $\Pi_i^{\mathrm{so}}$: optimal solo portfolio of farmer $i$.
    \item $\Pi_i^{\mathrm{co}}$: coalition portfolio adopted by farmer $i$.
    \item $\boldsymbol{\Pi}_S=\{\Pi_i^{\mathrm{co}}\}_{i\in S}$: coalition practice profile.

    \item $\alpha_{jk}$: pairwise interaction effect of practices $j$ and $k$ on sequestration.
    \item $\beta_{jk}$: pairwise interaction effect of practices $j$ and $k$ on operational cost.
    \item $\gamma_{jk}$: pairwise interaction effect of practices $j$ and $k$ on yield.

    \item $C_{\mathrm{MRV}}(S,\boldsymbol{\Pi}_S)$: MRV cost of coalition $S$.
    \item $C_{\mathrm{T}}(S)$: transaction cost of coalition $S$.
    \item $\mathrm{Fix}_{\mathrm{MRV}}, \mathrm{Var}_{\mathrm{MRV}}$: fixed and variable MRV cost parameters.
    \item $\mathrm{Fix}_{\mathrm{T}}, \mathrm{Var}_{\mathrm{T}}$: fixed and variable transaction-cost parameters.
    \item $\delta \in (0,1)$: MRV scale exponent.

    \item $\tilde v(\{i\})$: optimized standalone net value of farmer $i$.
    \item $v(S)$: characteristic-function value of coalition $S$.
    \item $\hat v(\{i\})$: adjusted solo value used in allocation constraints.
    \item $\mathbf{x}=(x_i)_{i\in S}$: allocation vector for coalition $S$.
    \item $x_i$: share of allocable carbon-market surplus assigned to farmer $i$.
    \item $e(S,\mathbf{x})$: excess of coalition $S$ under allocation $\mathbf{x}$.
    \item $\varepsilon$: least-core relaxation parameter.
    \item $\phi_i(v)$: Shapley value of farmer $i$.
\end{itemize}

\section{Dataset}
\label{appendix:dataset}
This section documents the construction, calibration, and verification of the \textsc{CarbonFARM-Bench} dataset used in the coalition formation experiments in the main paper, covering the numerical parameters, modeling assumptions, and data construction procedures adopted.

Section~\ref{sec:base} derives the three base parameters (\textbf{Carbon Mitigation Potential (CMP)}, \textbf{Net Operating Cost}, and \textbf{Yield Change}) for each of the eight carbon farming practices. Section~\ref{sec:interactions} explains how combining two practices can produce effects that are different from just adding them up, and how each pair is classified as having a positive, negative, or negligible interaction on carbon, cost, and yield. Section~\ref{sec:mrv} documents the MRV and transaction costs. Section~\ref{sec:data_farmers} explains the synthesis of farm sizes and operating budgets for the farmer groups. 

\subsection{Carbon Practices}
We consider the following eight carbon practices for our study. These practices are a combination of rice-specific and crop-agnostic practices. It should be noted that this list is agnostic and can be replaced with any set of practices.

\begin{itemize}    
\item Crop Residue Retention (CRR)
    \item Reduced / Minimum Tillage (RMT)
    \item Alternate Wetting and Drying (AWD)
    \item System of Rice Intensification (SRI)
    \item Direct Seeded Rice (DSR)
    \item Integrated Nutrient Management (INM)
    \item Manure Application (MA)
    \item Biochar Application (BA)
\end{itemize}

Each practice $j$ is characterized by three literature-derived base parameters: a carbon mitigation potential $\mathrm{CMP}_j$ (tCO$_2$e ha$^{-1}$ yr$^{-1}$), a net incremental operating cost $\Delta OC_j$ (USD ha$^{-1}$, PPP-adjusted), and a percentage change in grain yield $Y_j$ relative to conventional cultivation. When practices are adopted jointly, the model additionally applies pairwise interaction terms $\alpha_{jk}$ (carbon), $\beta_{jk}$ (cost), and $\gamma_{jk}$ (yield), plus a binary hard incompatibility indicator $\Delta_{jk}$, so that joint outcomes need not equal the naive sum of individual effects. Coalition-level enrollment additionally incurs MRV and transaction costs that scale with enrolled area and farmer count.

\subsection{Base Parameter Derivation}
\label{sec:base}
\subsubsection{Carbon Mitigation Potential (CMP)}

Carbon Mitigation Potential (CMP) quantifies the annual climate mitigation benefit achieved by adopting a carbon farming practice relative to conventional continuously flooded, puddled transplanted rice cultivation. The mitigation benefit arises through two complementary mechanisms:

\begin{enumerate}
    \item reduction in greenhouse gas (GHG) emissions resulting from
    improved crop and water management; and
    \item additional soil organic carbon (SOC) stored in the soil.
\end{enumerate}

These two mitigation pathways are integrated into a single CO$_2$-equivalent metric, referred to as Carbon Mitigation Potential (CMP), which forms the basis for evaluating the climate benefit of each practice throughout this study.

The CMP of practice $j$ is therefore defined as
\begin{equation}
\mathrm{CMP}_j
=
E_0r_j
+
\frac{44}{12}\Delta SOC_j,
\label{eq:cmp_a}
\end{equation}

where

\begin{itemize}

\item $E_0 = 7.8$ tCO$_2$e ha$^{-1}$ yr$^{-1}$ denotes the baseline annual GHG emission from conventional rice cultivation \cite{qian2023greenhouse}. This values varies based on the crop.

\item $r_j$ denotes the fractional reduction in Global Warming Potential (GWP) attributable to practice $j$, where positive values represent lower GHG emissions relative to conventional management;

\item $\Delta SOC_j$ denotes the additional annual SOC sequestration (Mg C ha$^{-1}$ yr$^{-1}$) associated with practice $j$;

\item the factor $44/12 = 3.667$ converts stored carbon into its
CO$_2$ equivalent following the IPCC Guidelines
\citep{ipcc2019refinement}.

\end{itemize}
The formulation consists of two additive components corresponding to the mitigation pathways introduced above,
\[
\underbrace{E_0r_j}_{\text{Avoided GHG emissions}}
+
\underbrace{3.667\Delta SOC_j}_{\text{SOC sequestration}},
\]
allowing avoided emissions and soil carbon sequestration to be evaluated independently. This distinction is important because some practices increase SOC while simultaneously increasing methane emissions under flooded conditions. Combining these effects into a single empirical value would obscure the individual contribution of each mitigation pathway.

\begin{figure}[t]
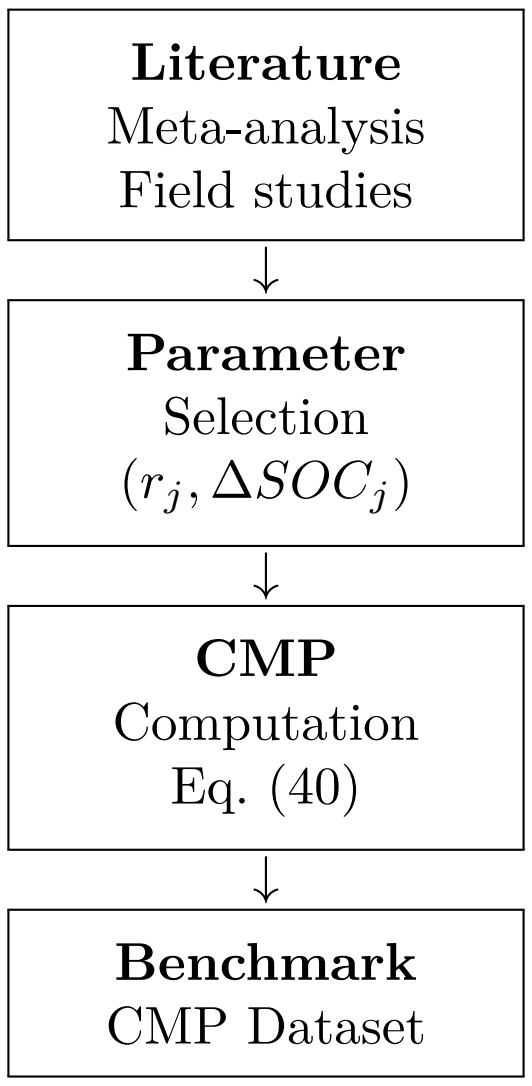

\centering
\setlength{\fboxsep}{6pt}

\fbox{\parbox{0.18\linewidth}{\centering
\textbf{Literature}\\
Meta-analysis\\
Field studies}}

$\downarrow$

\fbox{\parbox{0.18\linewidth}{\centering
\textbf{Parameter}\\
Selection\\
($r_j,\Delta SOC_j$)}}

$\downarrow$

\fbox{\parbox{0.18\linewidth}{\centering
\textbf{CMP}\\
Computation\\
Eq.~(\ref{eq:cmp_a})}}

$\downarrow$

\fbox{\parbox{0.18\linewidth}{\centering
\textbf{Benchmark}\\
CMP Dataset}}

\caption{Workflow for estimating Carbon Mitigation Potential (CMP) for
each carbon farming practice.}
\label{fig:cmpworkflow}
\end{figure}

Figure~\ref{fig:cmpworkflow} summarizes the procedure used to estimate CMP for each carbon farming practice. Literature-derived evidence is first used to determine the GWP reduction ($r_j$) and SOC sequestration ($\Delta SOC_j$) associated with each practice. These parameters are then substituted into Equation \ref{eq:cmp_a} to obtain the benchmark CMP values used throughout the optimization framework.

Biochar represents a special case within this framework because its carbon benefit originates primarily from a one-time addition of stable carbon rather than a continuous annual increase in native SOC. Consequently, $\Delta  OC_{\mathrm{BC}}$ is treated as an annualized stable-carbon credit calculated over a specified accounting horizon (Section~\ref{sec:biochar-cmp}).

\begin{table}[ht]
\centering
\small
\caption{Literature-informed carbon mitigation parameters used in the benchmark dataset.}
\vspace{1em}
\label{tab:cmp}
\begin{tabular}{cccc}
\toprule
\multirow{2}{*}{\textbf{Practice}} &
\multirow{2}{*}{\textbf{$r_j$}} &
\textbf{$\Delta SOC_j$} &
\textbf{CMP} \\
& & \textbf{Mg C ha$^{-1}$ yr$^{-1}$} & \textbf{tCO$_2$e ha$^{-1}$ yr$^{-1}$} \\
\midrule
Crop Residue Retention            & $-0.050$ & 0.27       & \textbf{0.60} \\
Reduced / Minimum Tillage         & 0.150    & 0.00       & \textbf{1.17} \\
Alternate Wetting and Drying      & 0.3684   & 0.00       & \textbf{2.87} \\
System of Rice Intensification    & 0.280    & 0.00       & \textbf{2.18} \\
Direct Seeded Rice                & 0.370    & 0.00       & \textbf{2.89} \\
Integrated Nutrient Management    & 0.000    & 0.50       & \textbf{1.83} \\
Manure Application                & 0.000    & 0.46       & \textbf{1.69} \\
Biochar Application               & 0.086    & 0.42$^{*}$ & \textbf{2.20} \\
\bottomrule
\vspace{1em}
\end{tabular}

\begin{minipage}{0.9\linewidth}
\footnotesize
$^{*}$For biochar, $\Delta SOC_j=0.42$ Mg C ha$^{-1}$ yr$^{-1}$ represents an annualized stable-carbon credit rather than a directly measured annual SOC change. It is calculated assuming 10 Mg biochar ha$^{-1}$, 60\% carbon content, 70\% stable carbon, and a 10-year accounting horizon.
\end{minipage}
\end{table}

Applying the above methodology yields the CMP parameter values summarized in Table~\ref{tab:cmp}. The scientific rationale, parameter selection, and detailed derivation for each practice are presented in the following subsections.

\subsubsubsection{\textbf{Crop Residue Retention}}

Crop residue retention returns organic biomass to agricultural soils, increasing soil organic carbon (SOC). Under flooded rice cultivation, however, decomposition of incorporated residues can stimulate methane production because of enhanced anaerobic microbial activity. The practice therefore affects both SOC sequestration and greenhouse gas emissions at the same time
\citep{lu2015,xia2024}.

\textit{Parameter Selection.}

Based on reported SOC sequestration rates, an annual SOC gain of
\[
\Delta SOC_{\mathrm{CRR}}
=
0.27~\mathrm{Mg~C~ha^{-1}~yr^{-1}}
\] is adopted.

To account for the increased methane emissions reported in flooded paddy systems, a conservative GWP penalty of 
\[ r_{\mathrm{CRR}} = -0.05 \] is assigned.

The GWP penalty is treated as a modeling assumption informed by the literature rather than a value directly reported by any single study.

\textit{CMP Calculation.}

Substituting the selected parameters into
Equation \ref{eq:cmp_a},

\[\mathrm{CMP}_{\mathrm{CRR}}=7.8\cdot(-0.05)+ 3.667\cdot(0.27) = 0.60. \]

\textit{Interpretation.}

Crop residue retention provides a positive overall mitigation benefit because increased SOC sequestration outweighs the assumed increase in methane emissions, giving the practice a moderate CMP value of 0.60 tCO$_2$e ha$^{-1}$ yr$^{-1}$.

\subsubsubsection{\textbf{Reduced / Minimum Tillage}}

Reduced or minimum tillage decreases soil disturbance and machinery operations, lowering fossil-fuel consumption and greenhouse gas emissions compared with conventional tillage. Literature syntheses consistently report reduced GWP under conservation tillage in rice-based production systems \citep{riceSynthesis2026}.

\textit{Parameter Selection.}

A representative GWP reduction of
 \[ r_{\mathrm{RMT}} = 0.15 \]

is adopted directly from the reported literature synthesis.

No independent SOC contribution is assigned,
 \[ \Delta SOC_{\mathrm{RMT}} = 0, \]
because SOC gains reported under conservation agriculture are often jointly attributed to reduced tillage and residue retention together.
Assigning an additional SOC benefit here would therefore lead to double counting.

\textit{CMP Calculation.}

Using Equation \ref{eq:cmp}, 
\[ \mathrm{CMP}_{\mathrm{RMT}}
= 7.8\cdot(0.15) = 1.17. \]

\textit{Interpretation.}

The mitigation benefit of reduced tillage arises entirely from avoided greenhouse gas emissions. Since no independent SOC credit is assigned, the resulting CMP reflects only the reduction in GWP.

\subsubsubsection{\textbf{Alternate Wetting and Drying (AWD)}}

Alternate Wetting and Drying (AWD) periodically drains paddy fields, interrupting the anaerobic soil conditions responsible for methane production. As a result, AWD is widely recognized as one of the most effective water-management practices for reducing methane emissions from irrigated rice cultivation.

\textit{Parameter Selection.}

A recent global meta-analysis reports approximately 43.2\% lower methane emissions and 36.8\% lower overall GWP under AWD relative to continuously flooded rice systems \citep{awd2024}. Accordingly,
\[ r_{\mathrm{AWD}} = 0.3684. \]

Because the principal mitigation pathway of AWD is methane reduction, no additional SOC sequestration benefit is assigned, 
\[ \Delta SOC_{\mathrm{AWD}} = 0. \]
 \textit{CMP Calculation.}

Applying Equation \ref{eq:cmp_a},

\[ \mathrm{CMP}_{\mathrm{AWD}} = 7.8\cdot(0.3684) = 2.87. \]

\textit{Interpretation.}

AWD achieves one of the highest CMP values among the evaluated practices because its mitigation benefit is dominated by substantial reductions in methane emissions rather than by SOC sequestration.
\subsubsubsection{\textbf{System of Rice Intensification (SRI)}}

The System of Rice Intensification (SRI) integrates improved crop, water, and soil management practices to enhance resource-use efficiency. Compared with conventional puddled transplanted rice, intermittent irrigation under SRI reduces the duration of anaerobic conditions, lowering methane emissions while maintaining or improving crop productivity.

\textit{Parameter Selection.}

Field experiments conducted in the Indo-Gangetic Plains report approximately 28\% lower net GWP under SRI compared with conventional rice cultivation \citep{jain2014}. Since $r_j$ represents the fractional reduction in GWP, the reported value is directly adopted as 
\[ r_{\mathrm{SRI}} = 0.28. \]

No independent SOC sequestration benefit is assigned,
\[ \Delta SOC_{\mathrm{SRI}} = 0, \]

because the reported mitigation primarily arises from improved water management rather than measurable increases in soil carbon storage.

\textit{CMP Calculation.}

Applying Equation \ref{eq:cmp_a},

\[ \mathrm{CMP}_{\mathrm{SRI}} = 7.8\cdot(0.28) = 2.18. \]

\textit{Interpretation.}

The mitigation benefit of SRI is driven almost entirely by reduced GHG emissions, so SRI exhibits one of the highest CMP values among the considered practices despite receiving no additional SOC credit.

\subsubsubsection{\textbf{Direct Seeded Rice (DSR)}}

Direct Seeded Rice (DSR) eliminates puddling and transplanting, reducing the duration of flooded conditions during crop establishment. This substantially suppresses methane production while also reducing labour and water requirements compared with conventional transplanted rice.

\textit{Parameter Selection.}

A recent meta-analysis reports approximately 69.9\% lower methane emissions and approximately 37\% lower overall GWP under DSR relative to puddled transplanted rice \citep{dsr2024}. 
Accordingly,
\[ r_{\mathrm{DSR}} = 0.37. \]
No separate SOC benefit is assigned,
\[ \Delta SOC_{\mathrm{DSR}} = 0, \]
because the reported climate benefit originates primarily from reduced methane emissions rather than enhanced soil carbon sequestration.

\textit{CMP Calculation.}

Using Equation \ref{eq:cmp_a},
\[\mathrm{CMP}_{\mathrm{DSR}}=7.8\cdot(0.37)=2.89.\]
\textit{Interpretation.}

Among all evaluated practices, DSR provides one of the largest climate mitigation benefits because its substantial reduction in methane emissions directly translates into a high CMP value.

\subsubsubsection{\textbf{Integrated Nutrient Management (INM)}}

Integrated Nutrient Management (INM) combines inorganic fertilizers with organic nutrient sources such as farmyard manure or compost. Long-term application of integrated nutrient management enhances soil organic matter inputs, improves soil health, and promotes gradual SOC accumulation.

\textit{Parameter Selection.}

Long-term rice-based experiments consistently report increased SOC under integrated nutrient management \citep{nayak2012,inm2023}. Based on these studies, an annual SOC sequestration rate of
\[\Delta SOC_{\mathrm{INM}}=0.50~\mathrm{Mg~C~ha^{-1}~yr^{-1}}\]
is adopted.

The net GWP response of INM varies considerably depending on the type of organic amendment, fertilizer application rate, and irrigation regime. No universal reduction in GWP is therefore imposed, and
\[r_{\mathrm{INM}}=0.\]
\textit{CMP Calculation.}

Substituting these values into Equation \ref{eq:cmp_a},
\[\mathrm{CMP}_{\mathrm{INM}}
=3.667\cdot(0.50)=1.83.\]

\textit{Interpretation.}

The mitigation benefit of INM is attributed entirely to increased SOC sequestration. Although no direct GHG reduction is assumed, long-term carbon storage results in a substantial CMP value.

\subsubsubsection{\textbf{Manure Application}}

Application of farmyard manure directly adds organic carbon to the soil, increasing soil organic matter and promoting long-term SOC sequestration. Numerous long-term rice--wheat experiments have demonstrated sustained increases in SOC following repeated manure application \citep{manure2013}.

\textit{Parameter Selection.}

Based on reported long-term sequestration rates, the annual SOC gain is taken as
\[\Delta SOC_{\mathrm{manure}}=0.46~\mathrm{Mg~C~ha^{-1}~yr^{-1}}.\]

Although manure improves soil fertility, it can also stimulate methane production under flooded rice cultivation. Since the overall GWP response depends strongly on field conditions and manure management, no independent GWP reduction is assumed:
\[r_{\mathrm{manure}}=0.\]
\textit{CMP Calculation.}

Applying Equation \ref{eq:cmp_a},
\[\mathrm{CMP}_{\mathrm{manure}}=3.667\cdot(0.46)=1.69.\]

\textit{Interpretation.}

The climate benefit of manure application is therefore derived solely from additional SOC sequestration. Any potential reduction in GHG emissions is conservatively excluded to avoid overstating the mitigation benefit.
\subsubsubsection{\textbf{Biochar Application}}

\label{sec:biochar-cmp}

Biochar contributes to climate change mitigation through two distinct mechanisms. First, biochar amendments can reduce greenhouse gas emissions from rice cultivation by improving soil physical and chemical properties, suppressing methane production. Second, biochar contains highly stable carbon that remains in the soil for many years, providing a long-term carbon storage benefit. Biochar is therefore the only practice considered in this study whose CMP includes both an avoided-emissions component and a stable-carbon sequestration component.

\textit{Parameter Selection.}

A recent global meta-analysis of rice production systems reports an average reduction of approximately 8.6\% in Global Warming Potential (GWP) following biochar application \citep{biochar2025}. Since $r_j$ represents the fractional reduction in GWP, the reported value is adopted directly as 
\[ r_{\mathrm{BC}} = 0.086. \]

The corresponding avoided-emissions component is therefore 
\[ E_0r_{\mathrm{BC}}
= 7.8\cdot(0.086) = 0.671~\mathrm{tCO_2e~ha^{-1}~yr^{-1}}. \]

Unlike the remaining practices, the carbon benefit of biochar results from a one-time application of stable carbon rather than a continuous annual increase in native SOC. Accordingly, the annual SOC credit is derived from representative biochar characteristics.

A representative application rate of
\[ A_{\mathrm{BC}} = 10~\mathrm{Mg~ha^{-1}} \] is assumed, together with
\[ f_C=0.60, \qquad f_{\mathrm{stable}}=0.70, \]  where $f_C$ denotes the carbon fraction of biochar and $f_{\mathrm{stable}}$ denotes the fraction of carbon assumed to remain stable over the accounting period.

The total stable carbon added by one application is therefore

\[
C_{\mathrm{stable}} = A_{\mathrm{BC}}\cdot f_C\cdot f_{\mathrm{stable}} = 10\cdot(0.60)\cdot(0.70) = 4.20~\mathrm{Mg~C~ha^{-1}}. \]
Because CMP is expressed on an annual basis, this stable-carbon stock is annualized over a 10-year accounting horizon,
\[ T_{\mathrm{BC}} = 10~\mathrm{yr}, \]
giving
\[ \Delta SOC_{\mathrm{BC}} = \frac{4.20}{10} = 0.42~\mathrm{Mg~C~ha^{-1}~yr^{-1}}. \]
The corresponding CO$_2$-equivalent contribution becomes
\[
3.667\cdot \Delta SOC_{\mathrm{BC}}
= 3.667\cdot(0.42)
= 1.540~\mathrm{tCO_2e~ha^{-1}~yr^{-1}}. \]

The assumed application rate, carbon fraction, stable-carbon fraction, and accounting horizon are explicit modeling  assumptions adopted to construct a representative benchmark dataset rather than universally applicable experimental constants.

\textit{CMP Calculation.}

Substituting both mitigation components into
Equation \ref{eq:cmp_a},

\[
\mathrm{CMP}_{\mathrm{BC}} = 7.8\cdot(0.086)+3.667\cdot(0.42)=0.671+1.540=2.2\]

\textit{Interpretation.}

Biochar achieves one of the highest CMP values among the evaluated practices because it combines two complementary mitigation pathways: direct reductions in GHG emissions and long-term storage of stable carbon in soil. Unlike the remaining practices, a substantial fraction of its mitigation benefit originates from annualized stable-carbon storage rather than changes in annual SOC dynamics alone.

\paragraph{Summary of CMP Estimation}

The CMP values presented above provide a consistent estimate of the annual climate mitigation potential of the eight carbon farming practices considered in this study. Although all practices are evaluated using the common framework of Equation \ref{eq:cmp}, the dominant mitigation pathway differs across practices. AWD, SRI, and DSR derive most of their mitigation benefit from reductions in methane emissions, whereas INM and manure application primarily contribute through enhanced SOC sequestration. Crop residue retention represents a balance between increased SOC storage and higher methane emissions, whereas biochar combines avoided emissions with long-term, stable carbon storage.

The resulting CMP dataset therefore represents a \emph{literature-informed benchmark} that integrates published experimental evidence with transparent modeling assumptions wherever direct estimates are unavailable. This consistent parameterization provides the basis for evaluating the climate benefits of alternative carbon farming portfolios within the cooperative optimization framework developed in this study.

\subsubsection{Net Operating Cost}

For practice $p$, net operating cost is the \emph{incremental} cost of adopting $p$ relative to conventional cultivation,
\begin{equation}
    \Delta OC_p = OC_p - OC_{\mathrm{conv}},
\end{equation}
where $\Delta OC_p < 0$ denotes a saving and $\Delta OC_p > 0$ denotes additional expenditure. The operating-cost boundary follows standard cost-of-cultivation accounting practice: hired and machine labour, land preparation, irrigation/pumping, fuel/energy, seed/nursery operations, fertilizer and manure, crop protection, and transport/handling of organic amendments. All local-currency figures are onverted to PPP-adjusted USD using a fixed exchange assumption of 1~USD$_{\mathrm{PPP}}$ = 25 local-currency units (LCU).

Table~\ref{tab:cost} summarizes the resulting values.

\begin{table}[ht]
\centering
\small
\caption{Net operating-cost assumptions relative to conventional cultivation.}
\vspace{1em}

\label{tab:cost}
\begin{tabular}{p{5.6cm}rr}
\toprule
\textbf{Practice} & \textbf{Net OC (LCU~/ha)} & \textbf{Net OC (USD/ha, PPP)} \\
\midrule
Crop Residue Retention        & $-395$    & $-15.80$ \\
Reduced / Minimum Tillage     & $-1{,}250$ & $-50.00$ \\
Alternate Wetting and Drying  & $-550$    & $-22.00$ \\
System of Rice Intensification & $+1{,}500$ & $+60.00$ \\
Direct Seeded Rice            & $0$        & $0.00$ \\
Integrated Nutrient Management & $+1{,}200$ & $+48.00$ \\
Manure Application            & $+3{,}000$ & $+120.00$ \\
Biochar Application           & $+20{,}000$ & $+800.00$ \\
\bottomrule
\end{tabular}
\end{table}

Each value is a \emph{literature-informed representative} parameter, not a directly reported empirical point estimate for the exact vector of practices, since no single study reports all eight incremental costs for a common location, season, and production system. The literature is instead used to establish which cost components are affected, the direction of the effect, and a plausible order of magnitude; a conservative representative value is then selected.
\begin{itemize}

\item\textbf{Crop Residue Retention} is assigned a small saving ($-$395~LCU/ha) reflecting avoided residue disposal net of any additional spreading/management requirement \citep{agronomy10122019}; the magnitude is deliberately conservative because it depends strongly on the residue-management baseline. 

\item\textbf{Reduced/Minimum Tillage} saves on machinery, fuel,
and tillage labour; against published cost-of-cultivation benchmarks of 12,342~LCU/ha machine labour and 31,770~LCU/ha hired labour \citep{kerala2013,agronomy9120886}, the assumed 1,250~LCU/ha saving is a conservative $\sim$10\% of the machine-labour line item.

\item\textbf{AWD} saves on irrigation/pumping, supported by reports of 23--33\% lower water use \citep{sahu2023, sudharani2020alternate}, positive economic returns from reduced pumping \citep{mote2020}, and up to $\sim$30\% irrigation savings \citep{richards2014}; only a small conservative saving ($-$550~LCU/ha) is monetized. 

\item\textbf{SRI} nets to a modest positive cost because assumed additional labour/management (transplanting, spacing, weeding) is taken to marginally exceed seed/nursery/irrigation savings \citep{TNAU_RiceEcosystem,pyasi2024comparative}.

\item\textbf{DSR} is assigned exactly 0~LCU/ha as a deliberately conservative assumption: reported labour/establishment savings are large (\citep{kumarDSR,Chaudhary2023DSR} reports $\sim$97\% labour and $\sim$80\% cost savings on the sowing/transplanting operation specifically) but are assumed to be offset by additional weed-management and herbicide costs, and no literature source directly supplies a net figure. 

\item\textbf{INM} and
\textbf{Manure} incur positive net costs reflecting procurement, transport, and application labour for organic inputs, offset only partially by fertilizer substitution \citep{thakur2023cost,manjappa2023}.

\item\textbf{Biochar} receives the largest incremental cost (+20,000~LCU/ha) because it additionally requires pyrolysis/production and transport \citep{Varma2025Biochar}; this figure is explicitly a scenario/calibration parameter rather than a value observed in the cited paddy-cultivation studies, and the report recommends sensitivity analysis around it.
\end{itemize}
\subsubsection{Yield Change}

For practice $j$, percentage yield change is
\begin{equation}
    Y_j = \frac{Y_{\mathrm{practice},j} - Y_{\mathrm{baseline}}}
    {Y_{\mathrm{baseline}}}\times 100,
\end{equation}
with $Y_j>0$ an increase, $Y_j<0$ a penalty, and $Y_j\approx0$ yield-neutral. Table~\ref{tab:yield} summarizes the values used.

\begin{table}[ht]
\centering
\small
\caption{Yield-change parameters for the eight final practices.}
\vspace{1em}

\label{tab:yield}
\begin{tabular}{p{5.4cm}cp{5.2cm}}
\toprule
\textbf{Practice} & \textbf{$Y_j$ (\%)} & \textbf{Basis} \\
\midrule
Crop Residue Retention & $+5.2$ & Rice-specific meta-analysis \citep{huang2013} \\
Reduced / Minimum Tillage & $+4.8$ & Conservative estimate based on South Asian rice systems reported by \citet{salahin2021, wang2024tillage} \\
Alternate Wetting and Drying & $0.0$ & Mild AWD maintains yield \citep{carrijo2017, GAO2024110075} \\
System of Rice Intensification & $+15.0$ & Regional field comparison \citep{reddy2013} \\
Direct Seeded Rice & $-12.0$ & Rice-specific meta-analysis \citep{xu2019} \\
Integrated Nutrient Management & $+1.4$ & Regional meta-analysis vs.\ 100\% NPK \citep{padbhushan2021} \\
Manure Application & $+3.9$ & Representative meta-analytic response \citep{manuremeta} \\
Biochar Application & $+10.7$ & Rice-specific meta-analysis \citep{liu2022} \\
\bottomrule
\end{tabular}
\end{table}

\begin{itemize}

\item\textbf{Crop Residue Retention:} a rice-specific meta-analysis of in-situ residue retention reports a 5.2\% yield increase \citep{huang2013}. 

\item \textbf{Reduced / Minimum Tillage:} A conservative estimate based on the 19--22\% yield gain reported for strip tillage in South Asian rice systems and 2.7\% in rice-rapseed cultivation \citep{salahin2021, wang2024tillage}.

\item\textbf{AWD:} a meta-analysis of 56 studies/528 comparisons found an average 5.4\% yield reduction overall, but this was driven by severe water stress; under properly managed \emph{mild} AWD (the practice represented here), yield was not significantly reduced, so $Y_{\mathrm{AWD}}=0\%$ is adopted \citep{carrijo2017}.

\item\textbf{SRI:} a regional field comparison reported 76.56 vs.\ 64.76 q/ha ($\approx$15\% increase) under SRI with recommended NPK \citep{reddy2013}; this conservative regional value is retained even though a broader 78-study, 27-country review reports larger average gains (24\% vs.\ recommended practice, 56\% vs.\ farmer practice) \citep{thakur2023}. 

\item\textbf{DSR:} a meta-analysis of 53 studies/440 paired comparisons reports a 12\% average yield penalty for DSR relative to transplanted rice \citep{xu2019}. 

\item\textbf{INM:} a regional meta-analysis reports a 1.4\% yield increase relative to the 100\% chemical-NPK comparison (the fully-fertilized baseline is used in preference to an unfertilized-control comparison, which would overstate the benefit) \citep{padbhushan2021}. 

\item\textbf{Manure:} synthesis evidence places partial manure-substitution yield gains at roughly 1.4--5.9\% and a broader grain-crop synthesis at roughly 3.3--3.9\%; a representative central value of $+3.9\%$ is adopted \citep{manuremeta}.

\item\textbf{Biochar:} a rice-specific meta-analysis reports a 10.73\% yield increase \citep{liu2022}, rounded to $+10.7\%$.
\end{itemize}

\subsection{Pairwise Interactions}
\label{sec:interactions}

Adopting two practices together need not produce outcomes equal to the sum of their individual effects. For example, SRI already bundles AWD-like irrigation and refined nutrient handling, so simply adding SRI's $+15\%$ yield term to Biochar's $+10.7\%$ would double-count the nutrient-management pathway both partly rely on.

We introduce pairwise interaction terms $\alpha_{jk}$ (carbon), $\beta_{jk}$ (cost), and $\gamma_{jk}$ (yield) for every unordered pair of the eight practices ($\binom{8}{2}=28$ pairs), plus a binary hard-incompatibility indicator $\Delta_{jk}\in\{0,1\}$. All four matrices are symmetric with zero diagonal.

\subsubsection{Categorical Classification and Sampling}
\label{sec:categories}

Each pair, in each dimension, is first assigned a category \textit{strong positive}, \textit{positive}, \textit{negative}, \textit{strongly negative}, or \textit{zero-like} based on mechanistic overlap or conflict (Section~\ref{sec:classification}). A numeric value is then drawn once, with a fixed random seed (42), from a category-specific normal distribution and clipped to a category-independent bound.

\begin{table}[ht]
\centering
\small
\caption{Sampling distributions (mean, sd) and clip bounds. Beta (cost) is shown in PPP-adjusted USD (divided by the PPP rate of 25 from the underlying INR distributions).}
\vspace{1em}

\label{tab:sampling}
\begin{tabular}{lccc}
\toprule
\textbf{Category} & $\alpha$ (carbon) & $\gamma$ (yield, \%) & $\beta$ (cost, USD/ha) \\
\midrule
strong\_pos  & 0.12 (0.03)   & 0.750 (0.375)  & $-48.0$ (10.0) \\
pos          & 0.08 (0.03)   & 0.375 (0.250)  & $-20.0$ (6.0) \\
neg          & $-0.15$ (0.04) & $-2.250$ (0.75) & $48.0$ (12.0) \\
neg\_strong  & $-0.25$ (0.05) & $-7.000$ (0.50) & $80.0$ (16.0) \\
zero\_like   & 0.00 (0.015)  & 0.000 (0.25)   & $0.0$ (4.0) \\
\midrule
\textbf{Clip bound} & $[-0.30,\,0.30]$ & $[-7.50,\,3.75]$ & $[-80.0,\,60.0]$ \\
\bottomrule
\end{tabular}
\end{table}

\subsubsection{Pairwise Classification}
\label{sec:classification}

Two principles govern the classification:

\begin{enumerate}
\item \textbf{A shared mechanism is penalized in every dimension it touches.} SRI already bundles AWD-like irrigation and refined nutrient handling; pairing it with INM, Manure, or Biochar re-exercises that same mechanism rather than adding an independent one, so these three pairs carry a strongly negative penalty ($S-$) in carbon, cost, \emph{and} yield  larger than an ordinary two-practice overlap, reflecting SRI's near-ceiling as a management package. INM and Manure likewise draw on the same organic-nutrient pathway, so this pair is negative in all three dimensions.
\item \textbf{Otherwise, carbon-adding practices are treated as complementary, not competing.} Crop Residue Retention, INM, Manure, and Biochar add carbon through distinct pathways (residue-C, organic/inorganic nutrients, stabilized biochar-C), so pairings among them  other than INM+Manure are positive or strongly positive in carbon and yield, not diminishing-returns.
\end{enumerate}

Table~\ref{tab:classification} lists every pair with a notable interaction. Pairs not listed default to zero-like.

\begin{table}[ht]
\centering
\small
\caption{Pairwise classification. \textit{S+} strongly positive, \textit{+} positive, \textit{$-$} negative, \textit{S$-$} strongly negative, \textit{--} zero-like (unclassified). Hard-excluded pairs ($\Delta=1$, Table~\ref{tab:hard-incompat}) are omitted here. Mechanism notes are given only where documented in the generator code; blank cells indicate a category assignment with no accompanying rationale in the code.}
\vspace{1em}
\label{tab:classification}
\begin{tabular}{lcccp{5cm}}
\toprule
\textbf{Pair} & $\alpha$ (C) & $\gamma$ (Y) & $\beta$ (cost) & \textbf{Mechanism} \\
\midrule
BA + CRR   & S+ & +  & -- & \\
BA + MA    & S+ & S+ & $-$ & Additive material/labour cost; stabilization and supply functions remain complementary \\
CRR + RMT  & S+ & +  & +  & Preserving carbon (RMT) and adding carbon (CRR) are complementary, not competing for the same pool \\
CRR + MA   & +  & +  & -- & \\
CRR + INM  & +  & S+ & -- & \\
CRR + DSR  & +  & $-$ & $-$ & Different mechanism per dimension: residue-carbon input (carbon) vs.\ drilling/weed interference (yield) vs.\ specialized equipment/weed control (cost) \\
AWD + DSR  & S+ & +  & S+ & \\
AWD + RMT  & -- & -- & +  & \\
AWD + BA   & +  & S+ & -- & \\
BA + INM   & +  & S+ & $-$ & Additive material/labour cost; stabilization and supply functions remain complementary \\
DSR + RMT  & $-$ & $-$ & S+ & Shared reduced-tillage operations save cost; water-management/establishment combination still costs yield and carbon \\
DSR + MA   & $-$ & $-$ & -- & \\
INM + MA   & $-$ & $-$ & $-$ & Same organic-nutrient-input pathway \\
INM + SRI  & S$-$ & S$-$ & $-$ & Nutrient-management overlap with SRI's own package (near-ceiling penalty) \\
MA + SRI   & S$-$ & S$-$ & $-$ & Nutrient-management overlap with SRI's own package (near-ceiling penalty) \\
BA + SRI   & S$-$ & S$-$ & $-$ & Nutrient-management overlap with SRI's own package (near-ceiling penalty) \\
\bottomrule
\end{tabular}
\end{table}

\subsubsection{Hard Incompatibilities}
\label{sec:hard-incompat}
Four pairs are treated as hard-infeasible ($\Delta_{jk}=1$), all involving SRI (Table~\ref{tab:hard-incompat}).

\begin{table}[ht]
\centering
\small
\caption{Hard incompatibilities in the final model.}
\vspace{1em}

\label{tab:hard-incompat}
\begin{tabular}{p{2.6cm}p{9.5cm}}
\toprule
\textbf{Pair} & \textbf{Reason} \\
\midrule
DSR + SRI & Contradictory crop-establishment mechanisms (direct seeding vs.\ transplanting). \\
AWD + SRI & AWD's intermittent-irrigation water management is already embedded within the SRI package. \\
CRR + SRI & The only non-$\Delta$ pair negative in all three matrices simultaneously -- the same ``triple-negative'' fingerprint as the confirmed DSR+SRI exclusion. \\
RMT + SRI & Treated as hard-excluded in the current model, consistent with the other SRI incompatibilities above. \\
\bottomrule
\end{tabular}
\end{table}

Two further pairs are left as soft, implementable interactions with intentionally mixed signs across dimensions (Table~\ref{tab:mixed}): each has a plausible mixed-objective explanation, such as a real cost saving coexisting with a real yield/carbon penalty.

\begin{table}[ht]
\centering
\small
\caption{Pairs with intentionally different signs across interaction dimensions.}
\vspace{1em}

\label{tab:mixed}
\begin{tabular}{p{2.6cm}p{9.5cm}}
\toprule
\textbf{Pair} & \textbf{Rationale} \\
\midrule
RMT + DSR & Shared reduced-tillage operations produce a real cost saving; the weed and water-management/establishment combination still carries a yield and carbon penalty. \\
INM + BA, MA + BA & Additive material/labour cost, but the underlying mechanisms (stabilization/retention vs.\ nutrient supply) remain complementary, so carbon and yield stay positive. \\
CRR + DSR & Different mechanism per dimension: residue-carbon input (carbon, positive) vs.\ drilling/weed interference (yield, negative) vs.\ specialized equipment and weed control (cost, negative). \\
\bottomrule
\end{tabular}
\end{table}

\subsection{MRV and Transaction Cost Calibration}
\label{sec:mrv}

Coalition enrollment incurs Monitoring, Reporting, and Verification (MRV) costs in addition to the practice-level operating costs described earlier.
Following the carbon-farming MRV framework of \citet{vanderheyden2026monitoring}, we distinguish between (i) technical MRV costs associated with monitoring and quantifying carbon outcomes and (ii) transaction and administrative costs associated with certification, reporting, verification, and farmer coordination. Accordingly, for a coalition $S$ with enrolled area $A(S)$ (ha) and $N=|S|$ participating farmers, the certification cost is modeled as

\begin{equation}
C_{\mathrm{cert}}(S)=
\underbrace{
F_{\mathrm{MRV}}
+
V_{\mathrm{MRV}}\cdot A(S)^{\delta}
}_{\text{Technical MRV}}
+
\underbrace{
F_T+V_T\cdot N
}_{\text{Transaction Cost}}.
\end{equation}

The baseline parameterization used in the numerical experiments is

\begin{equation}
F_{\mathrm{MRV}}=\$2000,\qquad
V_{\mathrm{MRV}}=\$80/\mathrm{ha}^{0.70},\qquad
\delta=0.70,\qquad
F_T=\$1000,\qquad
V_T=\$10/\mathrm{farmer}.
\end{equation}
\begin{itemize}
\item\textbf{Fixed technical MRV cost}
($F_{\mathrm{MRV}}=\$2000$)
represents the project-level infrastructure required for carbon-credit certification, including MRV setup, monitoring system configuration, data management, model deployment, and technical coordination. The European Commission reports an indicative upfront MRV expenditure of approximately EUR~1000 for soil-carbon certification together with substantial recurring monitoring and verification costs over the project lifetime \citep{EC2022}. Since the proposed formulation represents the overall project-level technical MRV infrastructure rather than only the initial setup cost, a conservative baseline value of \$2000 is adopted.

\item\textbf{Variable technical MRV cost}
($V_{\mathrm{MRV}}=\$80/\mathrm{ha}^{0.70}$)
captures the area-dependent component of monitoring expenditure. \citep{vanderheyden2026monitoring} identify activity-data collection and soil sampling as the dominant monitoring cost drivers and demonstrate that monitoring costs increase with monitored area while exhibiting significant economies of scale. As the literature does not provide a coefficient directly corresponding to the functional form
$V_{\mathrm{MRV}}A^\delta$, the value $V_{\mathrm{MRV}}=80$ is adopted as a calibrated modeling parameter representing the order of magnitude of variable monitoring costs.

\item\textbf{Economies of scale}
are represented through the exponent $\delta=0.70$. This sub-linear specification reflects the empirical observation that average MRV costs decrease with increasing project size because several monitoring activities, such as image processing, reporting, verification, and technical support, can be shared across participating farms.

\item\textbf{Fixed transaction cost}
($F_T=\$1000$) represents project-level certification, reporting, quality assurance, and third-party verification activities. This value is consistent with the average third-party verification cost of approximately EUR~1250 per project reported by
\citet{vanderheyden2026monitoring}.

\item\textbf{Variable transaction cost}
($V_T=\$10$/farmer)
captures residual farmer-level administrative activities including documentation, communication, record maintenance, and coordination with the certification agency. Since the major monitoring activities are already represented within the technical MRV component, only the remaining administrative burden is included in the transaction cost. Accordingly, $V_T$ is treated as a calibrated modeling parameter rather
than a directly estimated empirical quantity.
\end{itemize}

The proposed decomposition is motivated by the empirical findings of \citet{vanderheyden2026monitoring}, who report that monitoring accounts for approximately 77\% of total MRV expenditure, while reporting and verificsation together contribute approximately 23\%. Accordingly, monitoring-related activities are represented through the technical MRV component, whereas reporting- and verification-related activities are represented through the transaction cost component. Because technical MRV depends primarily on certified area whereas transaction costs depend primarily on the number of participating farmers, the relative contribution of these two components varies naturally with coalition composition. Therefore, the reported 77:23 decomposition serves as the conceptual motivation for the proposed cost structure rather than as an exact calibration target.

\subsection{Farmer Dataset}
\label{sec:data_farmers}
A panel of $n = 15$ farmers is used for most of the experiments. Each
farmer $F_i$ is characterized by a farm size, $\mathrm{FS}_i$ (ha), and
an operating budget, $B_i$. Farm sizes are generated from a truncated
log-normal distribution \citep{lowder2016number},
$\log(\mathrm{FS}_i)\sim\mathcal{N}(-0.5,\,1.0^2)$, truncated to the
range $[0.10,\,2.00]$ ha \citep{rapsomanikis2015economic} to represent
smallholder farms. The generated values are rounded to two decimal
places. This parameterisation reflects the smallholder structure, placing most farmers in the 0.2--1.0~ha range while admitting a small fraction of holdings near 2~ha \citep{agcensus2015_phase1_2018}. Each farmer's budget is set as
\begin{equation}
    B_i = \text{FS}_i \times \bar{b} \times (1 + \epsilon_i),
    \qquad \epsilon_i \sim \mathcal{N}(0,\;0.20^2), \quad \epsilon_i \in [-0.20,\;+0.20],
\end{equation}
where $\bar{b} = \$140/ha$  is the base budget rate. The budget is thus positively correlated with farm size but introduces farmer-level heterogeneity in per-hectare capital availability, ensuring that budget constraints bind differentially across the panel.

\end{document}